\documentclass[10pt,a4paper]{article}

\usepackage{amssymb}
\usepackage{amsmath}
\usepackage[all]{xy}
\usepackage{mathrsfs}
\usepackage{pstricks}
\usepackage{pst-3d}
\usepackage{pst-node}
\usepackage{pst-fill}
\usepackage{pst-coil}
\usepackage{graphicx}
\usepackage{lscape}

\newcommand{\Real}{\text{Re}}                  
\newcommand{\ds}{\displaystyle}

\newcommand{\df}{\frac{}{}}                              

\newcommand{\w}{\wedge}                                  
\newcommand{\wt}[1]{\widetilde{#1}}                       
\newcommand{\wh}[1]{\widehat{#1}}                         
\newcommand{\Lie}{\mathcal{L}}                           
\newcommand{\diff}[2]{\frac{d{#1}}{d{#2}}}                  
\newcommand{\SOdiff}[2]{\frac{d^{2}{#1}}{d{{#2}}^{2}}}      
\newcommand{\pdiff}[2]{\frac{\partial{#1}}{\partial{#2}}}   
\newcommand{\conj}[1]{\overline{#1}}                      
\newcommand{\tensor}{\otimes}
\newcommand{\qquadand}{\qquad\text{and}\qquad}
\newcommand{\quadand}{\quad\text{and}\quad}

\newcommand{\man}[1]{{\cal #1}}

\newcommand{\real}{\mathbb{R}}

\newcommand{\M}{\man{M}}
\newcommand{\ep}[1]{\epsilon_{#1}}
\newcommand{\g}{{g}}
\newcommand{\ginv}{\g^{-1}}
\newcommand{\cc}{c_{0}}
\newcommand{\rd}{\text{d}}
\newcommand{\ri}{\text{i}}
\newcommand{\Exp}[1]{\text{exp}\!\left[ #1 \right]}
\newcommand{\Me}[1]{{\mathbf e}^{#1}}
\newcommand{\Mb}[1]{{\mathbf b}^{#1}}
\newcommand{\Md}[1]{{\mathbf d}^{#1}}
\newcommand{\Mh}[1]{{\mathbf h}^{#1}}
\newcommand{\Mm}[1]{{\mathbf m}^{#1}}
\newcommand{\Mp}[1]{{\mathbf p}^{#1}}

\newcommand{\MB}[1]{{\mathbf B}^{#1}}
\newcommand{\MD}[1]{{\mathbf D}^{#1}}

\newcommand{\MM}[1]{{\mathbf M}^{#1}}
\newcommand{\MP}[1]{{\mathbf P}^{#1}}
\newcommand{\V}{\mathbb{V}^U}
\newcommand{\Vt}{\wt{\mathbb{V}}^U}

\newcommand{\dotMB}[1]{\dot{\MB{}}\hspace{-0.1cm}\phantom{i}^{#1} }
\newcommand{\dotMD}[1]{\dot{\MD{}}\hspace{-0.1cm}\phantom{i}^{#1} }

\newcommand{\dotMM}[1]{\dot{\MM{}}\hspace{-0.1cm}\phantom{i}^{#1} }
\newcommand{\dotMP}[1]{\dot{\MP{}}\hspace{-0.1cm}\phantom{i}^{#1} }
\newcommand{\J}[1]{\man{J}^{U}_{#1}}
\newcommand{\Ai}[1]{\text{Ai}\!\left[ #1 \right]}
\newcommand{\Bi}[1]{\text{Bi}\!\left[ #1 \right]}
\newcommand{\DAi}[1]{\text{Ai}'\!\left[ #1 \right]}
\newcommand{\DBi}[1]{\text{Bi}'\!\left[ #1 \right]}
\newcommand{\sgn}[1]{\text{sgn}(#1)}
\newcommand{\kappaa}{{\cal K}}

\newcommand{\tauK}[1]{\tau_{K}^{#1}}
\newcommand{\JUK}[1]{J_{K}^{U#1}}
\newcommand{\rhoUK}[1]{\rho_{K}^{U#1}}
\newcommand{\PD}[1]{\partial_{#1}}
\newcommand{\sd}{\underline{d}}               
\newcommand{\TA}[1]{\langle \; #1 \; \rangle} 

\newcommand{\RU}[1]{\rho^{U}_{#1}}
\newcommand{\dotRU}[1]{\dot{\rho}^{\,U}_{#1}}
\newcommand{\hatRU}[1]{\wh{\rho}^{\,U}_{#1}}

 \newcommand{\BY}[1]{\textsc{#1},}                                 
 \newcommand{\atque}{{\normalfont and\;}}                       
 \newcommand{\TITLE}[1]{\textit{#1},}                              
 \newcommand{\IN}[4]{\textit{#1}, \textbf{#2}  (#3)  #4}     
\newcommand{\SEM}{stress-energy-momentum }
\newcommand{\EM}{electromagnetic }
\newcommand{\bfr}{{\mathbf r}}

\newcommand{\Evec}{{\mathbf v}}
\newcommand{\Ewec}{{\mathbf w}}

\newcommand{\inn}{{\it in}}

\newcommand{\calV}{{\cal V}}
\newcommand{\dnmm}{ \wh{\mu}_{r}^{2}\man{N}^2 + \mu_{r}^{2}\wh{\man{N}}^{2}}
\newcommand{\Id}{I\!d}

\newcommand{\kk}{{\mathbf k}}
\newcommand{\Y}{{\kk,\omega}}
\newcommand{\FA}[1]{\check{\alpha}^{U}_{#1}(\Y)}
\newcommand{\Fee}[2]{\check{\mathbf e}^{U{#1} #2}\!(\Y)}   
\newcommand{\Fbb}[2]{\check{\mathbf b}^{U#1 #2}\!(\Y)}
\newcommand{\Fdd}[2]{\check{\mathbf d}^{U#1 #2}\!(\Y)}
\newcommand{\Fhh}[2]{\check{\mathbf h}^{U#1 #2}\!(\Y)}
\newcommand{\FBB}[2]{\check{\mathbf B}^{U#1 #2}\!(\Y)}   
\newcommand{\FDD}[2]{\check{\mathbf D}^{U#1 #2}\!(\Y)}
\newcommand{\KKK}{{\mathbf K}}
\newcommand{\Xde}{{\mathbf \zeta}^{de}(\Y)}
\newcommand{\Xdb}{{\mathbf \zeta}^{db}(\Y)}
\newcommand{\Xhe}{{\mathbf \zeta}^{he}(\Y)}
\newcommand{\Xhb}{{\mathbf \zeta}^{hb}(\Y)}
\newcommand{\Xdedag}{{\mathbf \zeta}^{de\,\dagger}(\Y)}
\newcommand{\Xdbdag}{{\mathbf \zeta}^{db\,\dagger}(\Y)}

\newcommand{\Xhbdag}{{\mathbf \zeta}^{hb\,\dagger}(\Y)}
\newcommand{\XD}{{\man{D}}(\Y)}
\newcommand{\XDD}{{\man{D}}(\Y)}

\newcommand{\Fnnr}{{\mathbf {n}}^{r}(\Y)}
\newcommand{\Fnnnr}{{\mathbf {n}}^{r}}
\newcommand{\FFee}[1]{\check{e}^{#1}(\Y)}
\newcommand{\Zde}{{\zeta}^{de}}

\newcommand{\Zhb}{{\zeta}^{hb}}
\newcommand{\eqq}{\;\;=\;\;}
\newcommand{\ESM}{appendix}

\begin{document}
\title{\textbf{The Electrodynamics of Inhomogeneous Rotating Media and the Abraham and Minkowski Tensors II: Applications} }
\date{}
\author{Shin-itiro Goto, Robin W. Tucker and Timothy J. Walton \\
Department of Physics, Lancaster University, Lancaster and \\
The Cockcroft Institute, Keckwick Lane, Daresbury, UK.}

\maketitle

\begin{abstract}{Electrodynamics, Continuum Mechanics, Constitutive Theory, Relativity}
Applications of the covariant theory of drive-forms are considered
for a class of perfectly insulating media. The distinction between
the notions of `classical photons' in homogeneous bounded and
unbounded stationary media and in stationary unbounded
magneto-electric media is pointed out in the context of the
Abraham, Minkowski and symmetrized Minkowski \EM \SEM  tensors.
Such notions have led to intense debate about the role of these
(and other) tensors in describing  \EM interactions in moving
media. In order to address some of these issues for material
subject to  the Minkowski constitutive relations,  the propagation
of harmonic waves through homogeneous and inhomogeneous, isotropic
plane-faced slabs at rest is first considered. To motivate the
subsequent analysis on accelerating media two classes of \EM modes
that solve Maxwell's equations  for uniformly rotating {\it
homogeneous} polarizable media are enumerated. Finally it is shown
that, under the influence of an incident monochromatic, circularly
polarized, plane \EM wave, the Abraham and symmetrized Minkowski
tensors induce different time-averaged torques on a uniformly
rotating {\it materially inhomogeneous} dielectric cylinder. We
suggest that this observation may offer new avenues to explore
experimentally the covariant
electrodynamics of more general accelerating media.\\
\end{abstract}

\section{Introduction}\label{Introduction}
This is paper II of a series of two papers. In paper I it was
shown how the notion of a Killing vector field on spacetime could
be used, together with a divergence-less total \SEM tensor for a
material continuum interacting with the \EM field, in order to
establish the dynamical classical evolution of the medium. It was
emphasized that the precise form  of an  electrodynamic force or
torque depends on the nature of the decomposition of the total
\SEM into parts describing its \EM interaction with the medium and
the \EM constitutive relations for the macroscopic Maxwell
equations. The motivation for this approach was to establish
whether such quantities could be used to discriminate by
experiment  between a number of \EM \SEM tensors that have been
proposed in the past to describe the \EM interaction in dielectric
media and  to provide a comprehensive framework for the analysis
of electrodynamic problems in accelerating media. In this paper,
we adopt the conventions of paper I and apply the formulation to
the computation of a time-averaged \EM torque produced in
stationary and rotating media by an incident plane harmonic \EM
wave using the \EM \SEM tensors proposed by Abraham and Minkowski.
The first sections demonstrate how the notion of the classical
`photon' in stationary media can be formulated in this framework.
It is emphasized that the linear momentum of such photons not only
depends on the choice of  an \EM \SEM  tensor in the medium but
whether the medium has an interface with the vacuum or anisotropic
and dispersive properties. It is then shown how the existence of a
family of transverse \EM modes satisfying Maxwell's macroscopic
equations subject to the Minkowski constitutive relations for a
uniformly rotating homogeneous insulating medium can be used to
solve the boundary value problem for a plane harmonic \EM wave
incident on a homogeneous rotating slab.  That the time-averaged
\EM torque  on such a slab is the same for the Abraham and
Minkowski \SEM tensors motivates an analysis of  torques on
rotating {\it inhomogeneous} cylinders. In this case we
demonstrate that the time-averaged torques are significantly
different.

\section{Classical `Photons' in Stationary Homogeneous Bounded and Unbounded Isotropic
Media}\label{PhotonSect}
Although the quantization of the \EM field in a bounded
polarizable medium is non-trivial the notion of the classical
`photon' has been used to highlight the different
predictions concerning the linear momentum of light obtained by
adopting different \EM \SEM tensors in such a medium. These
notions can be defined in terms of time-harmonic classical \EM
field configurations in homogeneous isotropic stationary media
and the time-averages of their energy and linear momentum in a
fixed spatial volume $\calV$.

If a scalar field  $A(\bfr,t)$ depends on time $t$, its
time-average over any time interval $T$ is
\begin{eqnarray*}
    \TA{A}(\bfr) &\equiv& \frac{1}{T}\int_{0}^{T} A(\bfr,t)\, \rd t.
\end{eqnarray*}
Furthermore, if $\mathbf{A}(\bfr,t)$ is a real spatial $p$-form,
$\mathbf{B}(\bfr,t)$ a real spatial $q$-form in any frame $U$
and
\begin{eqnarray*}
    \mathbf{A}(\bfr,t) \eqq \Real\left( \man{A}(\bfr)\,\Exp{-\ri \omega t} \right),  \quad \mathbf{B}(\bfr,t) \eqq \Real\left( \man{B}(\bfr)\,\Exp{-\ri\omega t}\right),
\end{eqnarray*}
where $\man{A}(\bfr)$ and $\man{B}(\bfr) $ are complex spatial
forms with complex conjugates $\conj{\man{A}}(\bfr),\, \conj{\man{B}}(\bfr) $, then
\begin{eqnarray*}
    \mathbf{A} \w \mathbf{B} &=& \frac{1}{2}\Real\left(\df \man{A}\w\man{B}\,\Exp{-2\ri\omega t}\right) + \frac{1}{2}\Real(\man{A}\w\conj{\man{B}}).
\end{eqnarray*}
Hence
\begin{eqnarray}\label{TA2F}
    \TA{\mathbf{A} \w \mathbf{B}}(\bfr) &=& \frac{1}{2}\,\Real(\man{A}(\bfr)\w\conj{\man{B}}(\bfr)),
\end{eqnarray}
if $T=\frac{2\pi}{\omega}$. For any bounded spatial $3$-form
$\alpha_{t}$ in frame $U$ with {\it arbitrary} time variation, its
time-average $\TA{\alpha}$ over any finite interval of time $T$ is
the 3-form
\begin{eqnarray}
    \TA{\alpha} &=& \frac{1}{T}\int_{0}^{T}\, \alpha_{t}\,\rd t.
\end{eqnarray}
It follows immediately that if  $\alpha_{t}$ is time-periodic, but not necessarily harmonic,  with
period $T$ (i.e. $\alpha_{t}=\alpha_{t+T}$) then
$\TA{\dot{\alpha}}=0$.

If $K$ is a {\it spacelike translational} Killing vector field the
time-averaged linear momentum associated with a drive-form having
orthogonal components $\JUK{},\rhoUK{}$ in a volume $\calV$
relative to $U$ is
\begin{eqnarray}
    \man{P}_{K}^{U}[{\calV}] &=& \frac{1}{\cc}\,\int_{\calV} \, \TA{\rho_{K}^{U}} .
\end{eqnarray}
Similarly since $U$ is a {\it timelike translational} Killing vector
field in Minkowski spacetime, the time-averaged energy associated
with the same drive-form in a volume $\calV$ is
\begin{eqnarray}
    \man{E}_{K}^{U}[\calV] &=& \int_{\calV} \, \TA{\rho_{U}^{U}} .
\end{eqnarray}
This classical energy can be parcelled into $N$ `energy quanta',
each of which corresponds to that carried by a harmonic plane-wave
quantum with energy $\hbar\omega$  {\it in the vacuum}:
\begin{eqnarray}
    \man{E}_{K}^{U}[\calV] &=& N \hbar\omega .
\end{eqnarray}
One then defines the time-averaged $K$-component of linear momentum
associated with a classical \lq\lq photon'' in a volume
$\calV$ of the medium to be
\begin{eqnarray}\label{PhotonMom}
    p^{U}_{K} &=&  \frac{\man{P}_{K}^{U}[\calV]}{N} = \frac{\hbar\omega}{\cc} \left( \frac{\int_{\calV}\, \TA{\rhoUK{}}}{\int_{\calV} \, \TA{\rho_{U}^{U}}}\right).
\end{eqnarray}
Different choices of \SEM tensor for the \EM field with the {\it
same} constitutive relation for the medium will in general give
different values for $p_{K}^{U}$. In particular for {\it plane
harmonic waves} propagating in a simple {\it unbounded}
non-accelerating homogeneous, isotropic medium (described by relative permittivity $\ep{r}$ and relative permeability $\mu_{r}$) in a direction
aligned with $K$, the non-symmetric Minkowski \EM \SEM tensor
yields
\begin{eqnarray}\label{MinkDiPhotMom}
     p_{K}^{U,M}  &=& \frac{\man{N}\hbar\omega}{\cc},
\end{eqnarray}
where $\man{N}=\sqrt{\ep{r}\mu_{r}}$ is the refractive index of the medium, while that calculated from the Abraham tensor yields
\begin{eqnarray}\label{AbrDiPhotMom}
     p_{K}^{U,AB}  &=& \frac{\hbar\omega}{\cc\man{N}}.
\end{eqnarray}
Furthermore from the symmetrized Minkowski tensor one finds
\begin{eqnarray}\label{SMDiPhotMom}
    p_{K}^{U,SM}  &=& \frac{1}{2}\left( p_{K}^{U,M} + p_{K}^{U,AB} \right) = \frac{\hbar\omega}{2\cc}\left(\frac{\man{N}^{2} +1}{\man{N}} \right).
\end{eqnarray}
The underlying origin for these distinctions is that (see tables
in Appendix A of paper I) although the contributions to the energy
density $\rho_{U}^{U}$ is the same for the non-symmetric Minkowski
tensor $T^{M}$, the symmetrized Minkowski tensor $T^{SM}$ and the
Abraham tensor  $T^{AB}$ (since the medium acceleration $A=0$) the
contributions to the corresponding momentum densities $\rhoUK{}$
are different.

However even if the medium is stationary, homogeneous and
isotropic the presence of boundaries can modify these results.
Given that all physical media do have interfaces with other media
including the vacuum the role of these interfaces between  media
with different constitutive properties is relevant in discussing
physical phenomena. To illustrate this point consider a simple
medium composed of a  stationary plane faced slab of arbitrary
finite thickness and constant relative  permittivity $\ep{r}$ and
relative permeability $\mu_{r}$. If a harmonic circularly
polarized plane wave is incident from the left vacuum half-space
normally on one plane face it will be partially reflected and
transmitted. Suppose it propagates through the slab into a
half-space composed of a different stationary simple homogeneous
isotropic medium but with constant relative permittivity
$\wh{\epsilon}_{r}$ and relative permeability $\wh{\mu}_{r}$. One can
readily compute the \EM fields in each region from the \EM
junction conditions and hence the different classical photon
momenta in the three different media. If the slab has its plane
normals parallel to the direction of propagation $K=\pdiff{}{z}$,
one finds for the different classical photon momenta in the slab:
\begin{eqnarray}
    p_{K}^{U,SM} &=& \frac{\hbar\omega}{\cc}\frac{\wh{\man{N}}\mu_{r}\wh{\mu}_{r}\,(1+\man{N}^2)}{\dnmm}, \quad
    p_{K}^{U,M} \eqq \frac{\hbar\omega}{\cc}\frac{2\,\man{N}^{2}\wh{\man{N}} \mu_{r} \wh{\mu}_{r}}{\dnmm} \\
    p_{K}^{U,AB} &=& \frac{\hbar\omega}{\cc}\frac{2\,\mu_{r}\wh{\mu}_{r}\wh{\man{N}}}{\dnmm}
\end{eqnarray}
where $\wh{\man{N}}^2=\wh{\epsilon}_{r}\wh{\mu}_{r}$. Each of
these momenta in the bounded slab depends on the  properties of
the medium outside the slab. Furthermore if the right-hand half
space is taken to be the vacuum ($\wh{\man{N}} = 1, \wh{\mu}_{r} =
1$) they are different from the momenta above for classical
photons in homogeneous {\it unbounded} media.

Analogous results can be obtained by calculating the total
time-averaged classical angular momentum about some point in
$\calV$ using a spacelike rotational Killing vector generating
rotations about an arbitrary direction in space. Again the
momentum densities $\rhoUK{}$ derived from $T^{M}$ and $T^{AB}$
are different so the corresponding classical `photon' helicities
in a finite volume $\calV$ are different. We stress that such
`photon' momenta and `photon' helicities are strictly
classical notions calculated from time averages of harmonic plane
waves in a stationary medium. As such they are not subject to the
same conservation laws as genuine electromagnetic quanta in the
medium.\\

\section{Classical `Photons' in Stationary Unbounded Homogeneous Magneto-electric Media}
In dispersive media, constitutive relations between the real spatial
fields $\Me{U},\Mb{U},\Md{U},\Mh{U}$ are, in general, non-local in spacetime.
If the medium is {\it spatially homogenous}, so that it has no
preferred spatial origin, then in Minkowski spacetime it is
possible to Fourier transform the inertial components of these
fields with respect to space and time, and work with transformed
local constitutive relations.

For any spatial $1$-form $\alpha^{U}$ on spacetime
with inertial components $\alpha^{U}_{a}(\bfr,t)$, define their complex
valued Fourier transforms $\FA{a}$ by
\begin{eqnarray}\label{FT}
    \alpha^{U}_{a}(\bfr,t) &=& \int_{-\infty}^{\infty} \, d\omega \int_{-\infty}^{\infty} \, d\kk \, \FA{a} \, \Exp{\ri\kk \cdot \bfr - \ri\omega t},
\end{eqnarray}
where $\alpha^{U}(\bfr,t)=\alpha^{U}_{a}(\bfr,t)\,e^{a}$, $\FA{}=\FA{a}\,e^{a}$, $\FA{a} \in \real$
and $\kk\in\real^{3}$. Then the source-free macroscopic Maxwell system reduces
to
\begin{eqnarray}
    \label{F1} \KKK \w \Fee{}{} &=& \omega\FBB{}{}, \qquad \KKK \w \Fhh{}{} \eqq -\omega\FDD{}{}
\end{eqnarray}
where the real propagation wave 1-form $\KKK\equiv\kk\cdot d\bfr$.
The remaining transformed Maxwell equations $\KKK \w \FBB{}{}=0$
and $\KKK \w \FDD{}{}=0$ follow trivially from (\ref{F1}). It also
follows trivially that $\Fee{}{} \w \FBB{}{} = 0$ (i.e. $\Fee{}{}$
is perpendicular to $\Fbb{}{}$). Similarly, $\FBB{}{} \w \KKK=0$
and $\FDD{}{} \w \KKK=0$.

Assume that in any inertial frame $U$ the medium is described in
terms of the real magneto-electric ($1,1$) spatial tensors
$\Xde,\Xhb, \Xdb,\Xhe$ satisfying the symmetry conditions
\cite{DGT,DGT2}: 
\begin{eqnarray*}
    \Xdedag = \Xde \,,\quad \Xhbdag = \Xhb \quadand \Xdbdag = -\Xhe,
\end{eqnarray*}
where the adjoint $T^\dagger$ of any spacetime tensor
$T$ which maps $p$-forms to $p$-forms is defined by:
\begin{eqnarray}
    \alpha\w\star T(\beta) &=& \beta\w\star T^\dagger(\alpha) \qquad\text{for all spacetime $p$-forms }\alpha,\beta
\end{eqnarray}
and the dispersive magneto-electric constitutive relations in the
$U$ frame are defined by
\begin{eqnarray}
    \label{FCR1} \Fdd{}{} &=& \Xde(\Fee{}{}) + \Xdb(\Fbb{}{}) \\
    \label{FCR2} \Fhh{}{} &=& \Xhe(\Fee{}{}) + \Xhb(\Fbb{}{}).
\end{eqnarray}
These will (by convolution) give rise to non-local spacetime
constitutive relations.

Substituting (\ref{FCR1}) and (\ref{FCR2}) in (\ref{F1}) yields a
degenerate 1-form eigen-equation for $\Fee{}{}$:
\begin{eqnarray}
\begin{split}\label{FDR}
     & \omega^{2} \Xde(\Fee{}{}) + \omega\Xdb  \left( \#( \KKK \w \Fee{}{} ) \right) \\
     & + \omega\#\left( \KKK \w \Xhe(\Fee{}{}) \right) + \# \left(\KKK\w \Xhb\left(\#(\KKK\w\Fee{}{}) \right) \right) = 0.
\end{split}
\end{eqnarray}
The field $\Fbb{}{}$ then follows from (\ref{F1}), (up to a
scaling) and $\Fdd{}{},\Fhh{}{}$ from (\ref{FCR1}),(\ref{FCR2})
respectively. Equation (\ref{FDR}) may be written
\begin{eqnarray}\label{FD1}
    \XDD(\Fee{}{})&=& 0,
\end{eqnarray}
defining the $(1,1)$ tensor $\XDD$. For non-trivial solutions
$\Fee{}{}$, the determinant of the matrix $\XD$ representing
$\XDD$ must vanish:
\begin{eqnarray}\label{det}
    \det(\XD) &=& 0.
\end{eqnarray}
Note that, in general, the roots of this dispersion relation are
not invariant under the transformation $\KKK\to\,-\KKK$. If one
writes $\kk= \hat\kk |\kk|$ in terms of the Euclidean norm
$|\kk|$, and introduces the refractive index $\man{N}=|\kk|
\frac{\cc}{\omega}>0$ and $\wh{\kk}$ in place of $\kk$, then
solutions propagating in the direction described by $\wh{\kk}$
with angular frequency $\omega>0$ correspond to roots of
(\ref{det}) (labelled $r$) that may be expressed in the form
$\man{N}_{r}=\man{F}_{r}(\wh{\kk},\omega)$. Thus, there can be a
set of distinct characteristic waves each with its unique
refractive index that depends on the propagation direction
$\wh{\kk}$ and frequency $\omega$. When the characteristic
equation (\ref{det}) is a quadratic polynomial in $\man{N}^2$ and
has two distinct roots that describe two distinct propagating
modes for a given $\omega$, the medium is termed {\it
birefringent}. Roots $\man{N}^2_{r}$ such that
$\man{N}_{r}(\wh{\kk}, \omega) \neq \man{N}_{r}(-\wh{\kk},
\omega)$ imply that harmonic plane waves propagating in the
opposite directions $\pm\wh{\kk}$ have different wave speeds.

Each eigen-wave will have a uniquely defined polarization obtained
by solving the independent equations in (\ref{FD1}) for
$\Fee{,r}{}$, up to normalization. Since $\Fee{,r}{}$ is complex, it
is convenient to introduce the eigen-wave normalization by writing
\begin{eqnarray*}
    \Fee{,r}{} &=& \FFee{U,r}\, \Fnnr,
\end{eqnarray*}
in terms of the complex 0-form $\FFee{U,r}$ and complex
polarization 1-form $\Fnnr$, normalised to satisfy
\begin{eqnarray}
    \conj{\Fnnr}\w \#\Fnnr &=& \#1
\end{eqnarray}
for each $r$. If one applies $\#\conj{\Fee{,r}{}}\w\#$ to
(\ref{FDR}), making use of the symmetries between the real
magneto-electric tensors $\Xde,\Xdb,\Xhe,\Xhb$, and evaluates it
with the eigen-wave $\Fee{,r}{}$, one obtains the {\it real} 0-form
dispersion relation for the characteristic mode $r$:
\begin{eqnarray*}
    \omega^{2}  \# \left( \conj{\Fnnnr} \w \# \Xde(\Fnnnr) \right) &+& \omega\, \#\left( \conj{\Fnnnr} \w \#\Xdb \left( \#\,(\KKK \w  \Fnnnr) \right)\right) +  \\
    +   \omega \,\# \left( \conj{\Fnnnr} \w \KKK \w \Xhe(\Fnnnr) \right)  &+& \# \left(\conj{\Fnnnr} \w \KKK \w \Xhb \left( \#(\KKK \w \Fnnnr) \right)\right) = 0,
\end{eqnarray*}
where $\Fnnnr \equiv \Fnnr$ and
$\KKK=\frac{\omega}{\cc}\man{N}\wh{\kk}\cdot d\bfr$ in terms of
$\man{N}$ and $\wh{\kk}$.

For illustration, consider an
{\it unbounded}  magneto-electric material with
\begin{eqnarray}
    \label{XMS} \Xde &=& \zeta^{de}(\Y) \,\, \Id , \qquad \Xhb \eqq  \zeta^{hb}(\Y)\,\,  \Id,
\end{eqnarray}
in terms of the scalars $\zeta^{de}(\Y),\zeta^{hb}(\Y)$  and the rank-3 identity tensor $\Id$ in space. The
medium is oriented in the laboratory spatial basis
$\{\PD{x},\PD{y},\PD{z}\}$ so that $\Xdb$ takes the particular
form
\begin{eqnarray}
    \label{XdbMS} \Xdb &=& \beta_{1}(\Y) \,\,dx \tensor  \PD{y} + \beta_{2}(\Y) \, dy \tensor \PD{x},
\end{eqnarray}
in terms of the real scalars $\beta_{1}(\Y),\beta_{2}(\Y)$. The matrix
representing $\Xdb$ in the laboratory basis takes the form
\begin{eqnarray}\label{Xdbmatrix}
    [\Xdb] \equiv \left( \begin{array}{ccc}
                                            0 & \beta_{2}(\Y)  & 0 \\
                                            \beta_{1}(\Y) & 0 & 0  \\
                                            0 & 0 & 0 \\
                                        \end{array}\right).
\end{eqnarray}
It follows that\footnote{From the adjoint relation $\Xdbdag =
-\Xhe$.}
\begin{eqnarray}
    \label{XheMS} \Xhe &=& -\beta_{2}(\Y) \,\,dx \tensor \PD{y} - \beta_{1}(\Y) \,\, dy \tensor \PD{x}.
\end{eqnarray}
Consider a (complexified) harmonic plane wave propagating through
the medium along the $z$-axis  and polarized in the $y$-direction:
\begin{eqnarray}\label{MEplane}
    \Fee{,r}{} &=& \man{E}\Exp{\ri kz-\ri\omega t}dy
\end{eqnarray}
for some (real) constant $\man{E}$, angular frequency $\omega>0$
and wave number $k$. From (\ref{det}), the dispersion relation
associated with a polarized eigen-mode $\Fee{}{}$ is
\begin{eqnarray}\label{poly}
    \zeta^{de}(\Y)\omega^{2} - \zeta^{hb}(\Y)k^{2} - 2\beta_{2}(\Y) k\omega &=& 0,
\end{eqnarray}
describing propagation in a direction determined by
$\sgn{k}\,\PD{z}$ with phase speed $|\omega/k|$ depending on the
values of $\zeta^{de}(\Y),\zeta^{hb}(\Y)$ and $\beta_{2}(\Y)$.

Using (\ref{F1}) and (\ref{MEplane}) enables one to construct the
1-forms $\Fee{,r}{}$, $\Fbb{,r}{}$, $\Fdd{,r}{}$, $\Fhh{,r}{}$.
Then from (\ref{PhotonMom}), the classical photon momenta
associated with different \EM \SEM tensors (see Appendix A of
paper I) may be calculated. The different
 magneto-electric photon momenta associated with the
direction $K=\PD{z}$ are:
\begin{eqnarray*}
   p_{K,\pm}^{U,M}(\omega) &=& \hbar k_{\pm},  \qquad
   p_{K,\pm}^{U,AB}(\omega) \eqq  \frac{\hbar\,\omega^{2}}{\cc^{2}\,k_{\pm}} \\
   p_{K,\pm}^{U,SM}(\omega) &=& \frac{1}{2}\left( p_{K,\pm}^{U,M}(\omega) + p_{K,\pm}^{U,AB}(\omega)\right),
\end{eqnarray*}
where $k_{\pm}$ are the roots of
\begin{eqnarray*}
    k &=& \frac{\omega}{\Zhb(\Y)}\left(-\beta_{2}(\Y) \pm \sqrt{\beta^{2}_{2}(\Y) + \Zde(\Y)\Zhb(\Y) \;} \right)
\end{eqnarray*}
with $\kk=(0,0,k)$. These momenta are independent of
$\beta_{1}(\Y)$ owing to the choice of wave polarization. For
$\beta_{2}(\Y)=0$ one writes $\Zde(\Y)=\epsilon_0\epsilon_r(\Y)$,
$(\Zhb(\Y) )^{-1}=\mu_0\mu_r(\Y)$ and the magneto-electric photon
momenta reduce to the classical photon momenta in a stationary,
but {\it dispersive}, homogeneous, isotropic, unbounded
polarizable medium, (\ref{MinkDiPhotMom})-(\ref{SMDiPhotMom}),
with a frequency dependent refractive index. \\

\section{Transverse (TDB) Modes in Uniformly Rotating Homogeneous Media}\label{TBDsect}
In \S\ref{PhotonSect}, one used the result that plane harmonic \EM
waves can freely propagate in a simple homogeneous isotropic
non-dispersive non-accelerating unbounded polarizable medium. For
bounded media one expects that boundary conditions will put
constraints on the propagation characteristics. For inhomogeneous
bounded media single harmonic plane waves are no longer supported
and if the medium is accelerated (whether bounded or not) finding
solutions to Maxwell's equations in the medium becomes non-trivial
in general.

A significant difference between the \EM \SEM tensors advocated by
Minkowski and Abraham is that irrespective of the \EM constitutive
relation describing  the medium the Abraham tensor depends {\it
explicitly} on the bulk 4-velocity field of the medium. This in
turn implies that its divergence will depend {\it explicitly} on
the bulk 4-acceleration of the medium. This characteristic feature
is frame independent and gives rise, in general, to a non-trivial
coupling between the medium acceleration and the \EM field. This
interaction is of course absent for media at rest or moving with
constant linear velocity in any inertial frame. By contrast a
uniformly rotating medium should be sensitive in principle to such
an interaction. In the following we will attempt to calculate the
significance of this effect. If it is possible to measure the
torque on a uniformly rotating medium as a function of rotation
speed this should in principle discriminate between the two \EM
\SEM tensors.

To expedite this program it proves necessary to excite, by some
means, \EM fields in an electrically neutral rotating polarizable
medium. Since in practice one must deal with finite media and
fields outside the body one is confronted with a difficult problem
of \EM scattering from a bounded moving polarizable medium. To
circumvent this we shall approach the problem in terms of the
transmission of incident plane harmonic waves through a thin cylindrical
slab uniformly rotating about its axis of symmetry. If the radius
of the circular cylinder greatly exceeds its thickness it is
reasonable to neglect the boundary conditions on its rim.

Player \cite{PLAYER} 
was one of the first to explore the propagation of waves in a
uniformly rotating (unbounded) medium using the non-relativistic
Minkowski constitutive relations. He assumed a rigidly rotating
medium with angular speed $r\Omega \ll \cc$ for all points a
distance $r$ from the axis of rotation and concluded that in this
approximation a certain type of plane harmonic wave in the medium
was subject to a dispersion relation dependent on $\Omega$. Later
Gotte, Barnett and Padgett \cite{GBP}
argued that there should be a spectrum of such modes.

We look first for harmonic plane wave modes in a uniformly
rotating homogeneous uncharged medium by supposing that the
magnetic induction field $\Mb{U}$ and electric displacement field
$\Md{U}$ are transverse to the propagation of a wave (with wave
number $k$ and angular frequency $\omega$) along the axis of
rotation. In an inertial frame $U$ introduce the spatial
cylindrical coframe $\{e^{1}=dr, e^{2}=r\,d\,\theta, e^{3}=d\,z
\}$ in cylindrical polar coordinates $(r,\theta,z)$ centred at an
interface of the medium and, for some complex amplitudes $B(r),
D(r)$, assume that a pair of the complexified interior spatial
fields take the form
\begin{eqnarray}
    \label{MODES} \Mb{U} &=& B(r)\Exp{ \ri(m+1)\theta + \ri kz - \ri\omega t }(e^{1} + \ri e^{2})  \\
    \label{MODESS} \Md{U} &=& D(r)\Exp{ \ri(m+1)\theta + \ri kz - \ri\omega t }(e^{1} + \ri e^{2}) ,
\end{eqnarray}
where $m\in\mathbb{Z}$ and $\omega>0$. Such fields will be
referred to as circularly polarized TDB modes. For constant
$\Omega$ the spatial vector $\V=\Omega \pdiff{}{\theta}$ describes
a rigid rotation in $U$ about the $z$-axis in these coordinates
and will be adopted to describe the bulk motion of the
medium\footnote{With $\Omega$ constant a real unconstrained medium
would not remain unstressed in a strictly rigid state according to
Newtonian continuum mechanics. Furthermore we eschew all issues
associated with notions of relativistic rigidity  assuming that
they are peripheral to the main discussion here.}.
Then to first-order (in $\frac{\nu}{\cc}$) the constitutive
relations derived\footnote[1]{See \ESM for further details of this
calculation.} from (2.23) of paper I for a simple rotating
homogeneous medium determine the interior electric and magnetic
fields as
\begin{eqnarray*}
    \Me{U} &=& \Exp{ \ri(m+1)\theta + \ri kz - \ri\omega t }\left( \frac{D(r)}{\ep{0}\ep{r}}(e^{1} + \ri e^{2}) + \left(1 - \frac{1}{\ep{r}\mu_{r}}\right)B(r)\Omega r e^{3} \right)  \\
    \Mh{U} &=& \Exp{ \ri(m+1)\theta + \ri kz - \ri\omega t }\left( \frac{B(r)}{\mu_{0}\mu_{r}}(e^{1} + \ri e^{2}) -  \left(1 - \frac{1}{\ep{r}\mu_{r}}\right)D(r)\Omega r e^{3} \right).
\end{eqnarray*}
These fields must be compatible with the source free Maxwell
system for time harmonic fields\footnote[2]{For {\it homogeneous}
media it is only necessary to solve these coupled equations since
the other spatial Maxwell equations are then automatically
satisfied.}:
\begin{eqnarray*}
    \sd \Me{U} &=& \ri\omega \MB{U} , \qquad \sd \Mh{U} \eqq -\ri\omega \MD{U}.
\end{eqnarray*}
Substituting into the  first Maxwell equation above requires a
2-form to be zero (i.e. each component to vanish):
\begin{eqnarray}
    \nonumber    e^{1} \w e^{2}: \qquad \diff{D(r)}{r} &=& \frac{m D(r)}{r} \\
    \nonumber    e^{1} \w e^{3}: \qquad \diff{B(r)}{r} &=& -\frac{B(r)}{r} - \frac{\mu_{r}}{(1-\ep{r}\mu_{r})\Omega}\left( \ep{r}\omega \frac{B(r)}{r} + \frac{\ri k}{\ep{0}}\frac{D(r)}{r} \right) \\
    \label{Dsol} e^{2} \w e^{3}: \qquad \hspace{0.25cm} D(r) &=& \ri\ep{0}\ep{r}\left[\frac{\omega}{k} - \frac{(m+1)\Omega}{k}\left(1-\frac{1}{\ep{r}\mu_{r}}\right)\right] B(r).
\end{eqnarray}
This system has the solution
\begin{eqnarray*}
    D(r) &=& Ar^{m}, \\
    B(r) &=& -\frac{\ri k\mu_{r}A}{\ep{0}\left[(m+1)(1-\ep{r}\mu_{r})\Omega + \ep{r}\mu_{r}\omega \right]}\,r^{m}
\end{eqnarray*}
for some constant $A$.  Similarly the second Maxwell equation
above gives
\begin{eqnarray}
    \nonumber e^{1} \w e^{2}: \qquad \diff{B(r)}{r} &=& \frac{m B(r)}{r} \\
    \nonumber e^{1} \w e^{3}: \qquad \diff{D(r)}{r} &=& -\frac{D(r)}{r} - \frac{\ep{r}}{(1-\ep{r}\mu_{r})\Omega}\left( \mu_{r}\omega \frac{D(r)}{r} - \frac{\ri k}{\mu_{0}}\frac{B(r)}{r} \right) \\
    \label{Bsol} e^{2} \w e^{3}: \qquad \hspace{0.25cm} B(r) &=& -\ri\mu_{0}\mu_{r}\left[ \frac{\omega}{k}  - \frac{(m+1)\Omega}{k}\left(1-\frac{1}{\ep{r}\mu_{r}}\right) \right]D(r) .
\end{eqnarray}
with solution
\begin{eqnarray*}
    B(r) &=& Cr^{m}, \\
    D(r) &=& \frac{\ri k\ep{r}C}{\mu_{0}\left[(m+1)(1-\ep{r}\mu_{r})\Omega + \ep{r}\mu_{r}\omega \right]}\,r^{m},
\end{eqnarray*}
for some constant $C$. The solutions (\ref{Dsol}) and (\ref{Bsol})
are compatible provided
\begin{eqnarray}\label{TBDdisp}
    k^{2} &=& \frac{\man{N}^{2}}{\cc^{2}}\left[ \omega - (m+1)\Omega\left(1- \frac{1}{\ep{r}\mu_{r}}\right) \right]^{2}.
\end{eqnarray}
For real $\omega$ a root $k$ of this dispersion relation may
become complex for certain values of $\Omega$ (and describe
evanescent waves). Restricting to real $k$ roots, with real
$\Omega$ and real $\man{N}>0 $ such that
\begin{eqnarray*}
    \omega - \Omega\left(1 - \frac{1}{\ep{r}\mu_{r}}\right) &>& 0.
\end{eqnarray*}
yields propagating solutions with wave numbers as real roots of
the dispersion relation (\ref{TBDdisp}). These determine the two
possible directions of propagation of each TDB mode labelled by
$m$. Let
\begin{eqnarray}\label{kappa}
    k_{R,L}  &=& \eta_{R,L}\frac{\man{N}}{\cc}\left[ \omega - (m+1)\Omega\left(1- \frac{1}{\ep{r}\mu_{r}}\right) \right],
\end{eqnarray}
where $\eta_{R}=1$ and $\eta_{L}=-1$. Then $\sgn{k_{R}}>0$ and
$k_{R}$ denotes a right-moving wave while  $\sgn{k_{L}}<0$ and
$k_{L}$ denotes  a left-moving wave. Thus, for propagating TDB
waves satisfying the dispersion relation (\ref{TBDdisp})
\begin{eqnarray*}
    C_{R,L} &=& -\eta_{R,L}\frac{\ri\man{N}A_{m}^{+}(k_{R,L})}{\ep{0}\ep{r}\cc}.
\end{eqnarray*}
and
\begin{eqnarray*}
    B_{R,L}(r) &=& -\eta_{R,L}\frac{\ri\man{N}A_{m}^{+}(k_{R,L})}{\ep{0}\ep{r}\cc}r^{m} \\
    D(r) &=& A_{m}^{+}(k_{R,L})\,r^{m}
\end{eqnarray*}
for the arbitrary constant $A_{m}^{+}(k_{R,L})$. In this notation
the left and right propagating, left and right circularly polarized, TDB
modes in a uniformly rotating homogeneous dielectric cylinder can be written:
\begin{eqnarray*}
\begin{split}
    \Me{U,\pm}_{R,L} &= \mbox{\footnotesize $\frac{A_{m}^{\pm}(k_{R,L})}{\ep{0}\ep{r}}\,\Exp{ \ri(m+1)\theta + \ri k_{R,L}z - \ri\omega t }\left( e^{1} \pm \ri e^{2} - \ri \eta_{R,L}\left(1-\frac{1}{\ep{r}\mu_{r}} \right)\frac{\man{N}\Omega r}{\cc} e^{3} \right)r^{m}$}  \\
    \Mb{U,\pm}_{R,L} &= \mbox{\footnotesize $\eta_{R,L}\frac{A_m^{\pm}(k_{R,L})\mu_{0}\mu_{r}\cc}{\man{N}}\,\Exp{ \ri(m+1)\theta + \ri k_{R,L}z - \ri\omega t }(e^{2} \mp \ri e^{1})r^{m}$}  \\
    \Md{U,\pm}_{R,L} &= \mbox{\footnotesize $A_{m}^{\pm}(k_{k,L})\,\Exp{ \ri(m+1)\theta + \ri k_{R,L}z - \ri\omega t }(e^{1} \pm \ri e^{2})r^{m}$} \\
    \Mh{U,\pm}_{R,L} &= \mbox{\footnotesize $\frac{A_{m}^{\pm}(k_{R,L})\cc}{\man{N}}\,\Exp{ \ri(m+1)\theta + \ri k_{R,L}z - \ri\omega t }\left( \eta_{R,L}\left(e^{2} \mp \ri e^{1}\right) - \left(1-\frac{1}{\ep{r}\mu_{r}}\right)\frac{\man{N}\Omega r}{\cc}  e^{3} \right)r^{m}$}.
\end{split}
\end{eqnarray*}
If the axis $r=0$ is in the medium and there are no axial sources
then the integer $m\geq 0$.  All integers $m$ are permitted if the
medium is a shell with the rotation axis excluded. Fields with
$m>0$ are clearly unbounded if $r$ extends indefinitely. All modes
are eigen-forms of the rotation operator
$\frac{1}{\ri}\Lie_{\PD{\theta}}$ with eigenvalue $m+1$. The
interesting physical modes in an unbounded medium are the harmonic
plane waves with $m=0$. These are the waves first explored by
Player \cite{PLAYER}. 
One expects that these modes could be excited by a harmonic plane
wave normally incident on a rotating slab from the vacuum. In a
simple medium its refractive index $\man{N}$ is dispersion-free
and independent of the harmonic wave polarization.

There also exists a set of  circular polarized TEH modes
that can be calculated using an initial ansatz where the magnetic
field $\Mh{U}$ and electric field $\Me{U}$ are transverse to the
direction of propagation along the rotation axis:
\begin{eqnarray}
    \Me{U} &=& E(r)\Exp{ \ri(m+1)\theta + \ri kz - \ri\omega t }(e^{1} \pm \ri e^{2}) \label{MODES1} \\
    \Mh{U} &=& H(r)\Exp{ \ri(m+1)\theta + \ri kz - \ri\omega t }(e^{1} \pm \ri e^{2}) \label{MODES2}
\end{eqnarray}
with $m\in\mathbb{Z}$ as before. However if $\Omega\neq 0$ it
appears that there are no harmonic plane wave TEH modes among this set,
even for $m=0$, so we will not discuss them further here.

If a plane circularly polarized harmonic wave in the vacuum is
incident normally on a homogeneous uniformly rotating thin
cylindrical slab consisting of material satisfying a simple
constitutive relation (with $\ep{r}, \mu_{r}$ constant) then it is
straightforward to satisfy the \EM jump conditions at the slab
plane interfaces with left and right moving $m=0$, circularly
polarized, TDB modes in the medium (and a reflected and
transmitted circularly polarized, harmonic plane wave in the
vacuum). From equation (3.2) of paper I, one can use such a
solution to calculate the real instantaneous \EM $K$-drive on the
slab. Such a drive will fluctuate harmonically in time with
angular frequency a multiple of the incident frequency. For any
$\tauK{U,EM}$ the field in the slab is time-harmonic in the
laboratory frame $U$ and the time-average of
$\Lie_{U}\rho_{K}^{U,EM}$ is zero. Furthermore since
$J_{K}^{U,EM}$ is a real quadratic function of the time-harmonic
fields $\Me{U},\Mb{U},\Md{U},\Mh{U}$ in the medium, the
time-average of $J_{K}^{U,EM}$ is independent of the propagation
coordinate $z$. Hence from equation (3.2) of paper I, the
time-averaged total $K$-drive on a fixed volume $\calV$ of the
rotating  medium is:
\begin{eqnarray}\label{ROT_HOMOG}
    \conj{ f_{K}^{U,\inn,ext}[\calV] } &\equiv& -\int_{\PD{}\calV} \, \TA{ J_{K}^{U,\inn,EM} }.
\end{eqnarray}
The surface $\PD{}\calV $ consists of the two opposite interior
circular discs and the interior cylindrical rim. For all the
tensors in Appendix A of paper I, the contribution to the integral
from the latter is zero since the integrand in (\ref{ROT_HOMOG})
is zero when pulled back to the interior rim. Furthermore the
contributions to the integral from the two opposite interior faces
cancel since the time-averaged $\TA{ J_{K}^{U,\inn,EM} }$ is
independent of $z$. Hence one concludes that if a normally
incident plane harmonic vacuum wave excites a plane harmonic TDB
mode in such a simple (uniformly rotating or stationary) medium
the induced time-averaged $K$-drive will be zero. Thus one cannot
discriminate between Minkowski and Abraham \SEM tensors by
exciting plane TDB modes in a simple medium in this manner. One
alternative is to excite harmonic fields in a medium in which the
time-averaged $\TA{ J_{K}^{U,\inn,EM} }$ is not independent of
$z$. Then the contribution from opposite interior faces will not
necessarily vanish. The simplest strategy to implement this
requirement is to employ  fields in an inhomogeneous
dispersion-free medium in which the permittivity or permeability
varies with $z$ along the axis of the slab. Unfortunately the
simple TDB plane wave modes above are no longer solutions to the
Maxwell equations in such an inhomogeneous medium. Nevertheless it
will be shown below how the computation of the Maxwell fields in
such cases can lead to a method of discriminating between
different \EM \SEM tensors in uniformly rotating inhomogeneous
media.\\

\section{Plane Waves Incident on a Stationary Inhomogeneous Dielectric Slab}

The theory of Killing forms was exploited above in order to calculate the time-averaged harmonic \EM forces and torques on perfectly insulating homogeneous polarizable media described by the Minkowski constitutive relations. For stationary or non-relativistically uniformly rotating media it was found that the induced \EM drives did not discriminate between the \EM \SEM  tensors proposed by Minkowski and Abraham. If one stays with slabs excited by incident plane harmonic vacuum \EM waves it is natural to enquire about the influence of material inhomogeneities on this result.
However a polarized plane harmonic wave incident in the vacuum on a {\it stationary
inhomogeneous slab} of polarizable material will not in general
propagate in the medium as a polarized plane harmonic wave. To facilitate a
discussion in the next section of the behaviour of such an
incident wave on a {\it rotating  slab}, a solution to the problem of
fields in a stationary inhomogeneous slab will be considered in
this section.

\begin{figure}[h]
\begin{center}
    \includegraphics[width=13.5cm]{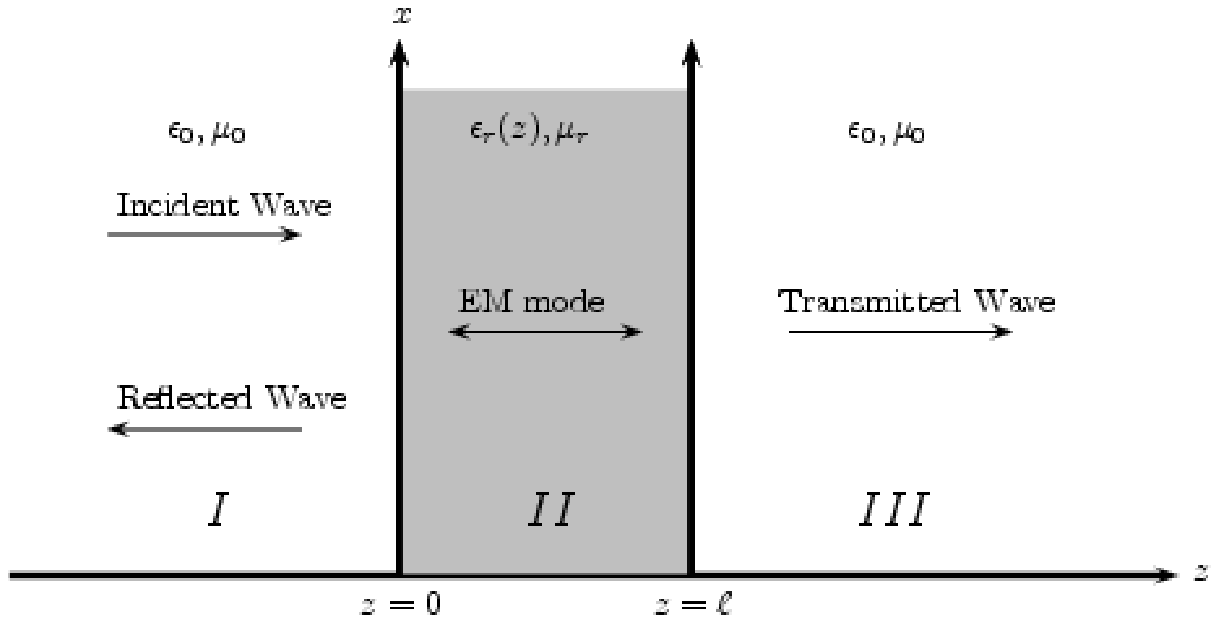}
\caption{Geometry of the inhomogeneous stationary dielectric slab
and the propagation direction of waves in the vacuum
regions.}\label{MagSlabDiag}
\end{center}
\end{figure}

The stationary slab is oriented in a Cartesian frame with
coordinates $\{x,y,z\}$ as indicated in figure~\ref{MagSlabDiag}.
Suppose a plane polarized plane wave monochromatic wave with
angular frequency $\omega>0$, propagating in the $z$-direction, is
incident on a slab of thickness $\ell$ and infinite extent in the
$x$ and $y$ directions.

Let the slab have relative permittivity $\ep{r}(z)$ and constant
relative permeability $\mu_{r}$. The slab has parallel interfaces
(with the vacuum) at $z=0$ and $z=\ell$ thereby distinguishing the
spatial region $z<0$ denoted $I$, $0<z<\ell$ denoted $II$ and
$z>\ell$ denoted $III$.

Let $F^{II}= d\,A^{II}$ in region $II$ with
\begin{eqnarray*}
    A^{II} &=& \man{A}_{0}\,\man{A}(z)\Exp{-\ri\omega t} dy
\end{eqnarray*}
for some (complex) function $\man{A}(z)$ and (real) constant
$\man{A}_{0}$ with physical dimensions  
(so that $\man{A}(z)$ is dimensionless). Then
\begin{eqnarray*}
    F^{II} = d\,A^{II} &=& \left(\diff{\man{A}(z)}{z}dz \w dy - \ri\omega\man{A}(z)dt \w dy\right)\man{A}_{0}\,\Exp{-\ri\omega t} .
\end{eqnarray*}
The medium is at rest relative to the inertial frame $U$ so is
assigned the 4-velocity $V=\frac{1}{\cc}\PD{t}$. Thus, from the
covariant dielectric constitutive relation (2.23) of paper I:
\begin{eqnarray*}
    G^{II} &=& \left(\frac{1}{\mu_{r}}\diff{\man{A}(z)}{z}dz \w dy - \ri\ep{r}(z)\omega\man{A}(z)dt \w dy\right)\man{A}_0\,\ep{0}\,\Exp{-\ri\omega t} .
\end{eqnarray*}
For an uncharged medium ($j=0$) the inhomogeneous Maxwell equation
(2.21) of paper I yields
\begin{eqnarray*}
    d\star G^{II} &=& -\left(\frac{1}{\mu_{r}}\SOdiff{\man{A}(z)}{z} + \frac{\omega^{2}\ep{r}(z)\man{A}(z)}{\cc^{2}}\right)\man{A}_{0}\,\cc\ep{0}\,\Exp{-\ri\omega t} dz \w dx \w dt = 0.
\end{eqnarray*}
If the relative permittivity is taken to be the function,
$\ep{r}(z)=\alpha + \frac{\beta z}{\ell}$ for some (dimensionless)
constants $\alpha,\beta$, this Schr\"{o}dinger-like differential
equation can be solved in terms of a basis of Airy functions of
the first and second kind (Ai and Bi respectively) to give
\begin{eqnarray*}
    \man{A}(z) &=& C_{1}\Ai{\kappaa(z)} + C_{2}\Bi{\kappaa(z)},
\end{eqnarray*}
where $C_{1},C_{2}\in\mathbb{C}$ and
\begin{eqnarray*}
    \kappaa(z) &=& -\left(\frac{\alpha \ell}{\beta} +  z\right)\left(\frac{\omega^{2}\beta\mu_{r}}{\ell \cc^{2}}\right)^{\frac{1}{3}}.
\end{eqnarray*}
Thus, for this  inhomogeneous dielectric with linearly-varying permittivity:
\begin{eqnarray}\label{InhomStatDiII}
    A^{II} &=& \man{A}_{0}\left( \df C_{1}\Ai{\kappaa(z)} + C_{2}\Bi{\kappaa(z)} \right)\Exp{-\ri\omega t} dy.
\end{eqnarray}
The dimensionless complex constants $C_{1},C_{2}$ can be
determined from two interface conditions. Consider vacuum
electromagnetic plane waves in region $I$ with (real) amplitude
$\man{A}_{0}$ and complex amplitude $E^{I}_{L}$, propagating in
the (positive and negative) $z$-directions respectively:
\begin{eqnarray}\label{InhomStatVacI}
    A^{I} &=& \man{A}_{0}\left(\df \Exp{\ri k^{I}_{R}z - \ri\omega t} + E^{I}_{L}\Exp{\ri k^{I}_{L}z - \ri\omega t}\right)dy,
\end{eqnarray}
where $k_{R}$ and $k_{L}$ denote distinct roots of the vacuum
dispersion relation
\begin{eqnarray}\label{VacDisp}
    k^{2} - \frac{\omega^{2}}{\cc^{2}} &=& 0
\end{eqnarray}
with $\sgn{k^{I}_{R}} > 0$ and $\sgn{k^{I}_{L}} < 0 $. Similarly,
a transmitted field in region $III$ is written
\begin{eqnarray}\label{InhomStatVacIII}
    A^{III} &=& \man{A}_{0}\,E^{III}_{R}\Exp{\ri k^{III}_{R}z - \ri\omega t}dy
\end{eqnarray}
for some complex amplitude $E^{III}_{R}$ and $\sgn{k^{III}_{R}}
> 0$. At each interface one must satisfy the interface conditions on a spacetime hypersurface\footnote{See \ESM for their formulation on spacetime.}. On the left interface $f=z$ and on the right interface
$f=z-\ell$. If $\Omega^{*}_{0}$ ($\Omega^{*}_{\ell}$) denote the
pull-back of forms to $z=0$ ($z=\ell$) respectively, the interface
boundary conditions become
\begin{eqnarray*}
    \Omega^{*}_{0}\left(F^{I} - F^{II}\right) &=& \Omega^{*}_{\ell}\left(F^{II} - F^{III}\right) = 0 \\
    \Omega^{*}_{0}\left(\star \,G^{I} - \star \,G^{II}\right) &=& \Omega^{*}_{\ell}\left(\star\, G^{II} - \star \,G^{III}\right) = 0.
\end{eqnarray*}
These yield a linear system of equations for the dimensionless
complex variables $E^{I}_{L},C_{1},C_{2},E^{III}_{R}$:\\

$\left(\begin{array}{cccc}
   \ds 0
& \Ai{\kappaa(\ell)} & \Bi{\kappaa(\ell)}
& \ds -\Exp{\frac{\ri\omega\ell}{\cc}}  \\
\ds 0 & \DAi{\kappaa(\ell)} & \DBi{\kappaa(\ell)}
& \ds   -\ri\mu_r\frac{\omega}{\cc}\Exp{\frac{\ri\omega\ell}{\cc}}\\
\ds \ri\mu_{r}\frac{\omega}{\cc} & \DAi{\kappaa(0)} &
\DBi{\kappaa(0)}
& \ds 0 \\
\ds -1 & \Ai{\kappaa(0)} & \Bi{\kappaa(0)} & \ds 0
\end{array}\right)
\left(\begin{array}{c}
   E^{I}_{L} \\
   C_{1} \\
   C_{2} \\
   E^{III}_{R}
\end{array}\right) =
\left(\begin{array}{c}
   \ds 0  \\
   \ds 0\\ 
   \ds \ri\mu_r\frac{\omega}{\cc} \\
   \ds 1
\end{array}\right) $ \\ \quad \\

where
\begin{eqnarray*}
    \DAi{\kappaa(a)} &\equiv& \left. \diff{}{z}\Ai{\kappaa(z)}\;\right|_{z=a}
\end{eqnarray*}
and similarly for $\DBi{\kappa(a)}$. The system admits a
 solution provided the $4\times 4$ complex matrix above
is non-singular.

\begin{figure}[h]
    \centering
    \includegraphics[scale=0.52,trim=0 0 200 0,clip=true]{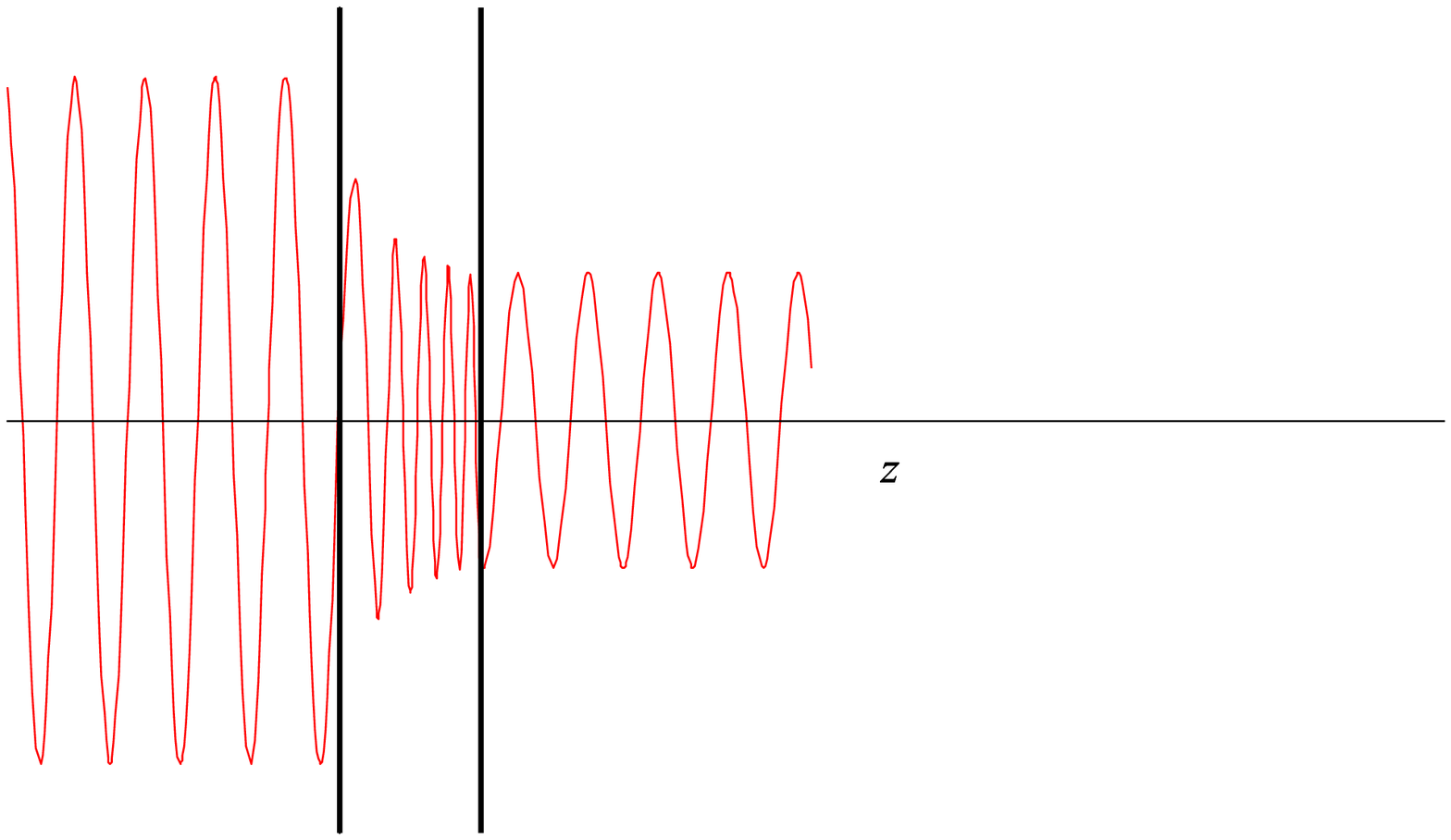}
    \caption{Electric field amplitude demonstrating transmission of a normally incident harmonic plane wave from the left  through a stationary linearly-inhomogeneous plane dielectric slab.}\label{electricamp}
\end{figure}

The solution is readily obtained but its complicated structure
will not be displayed here. It replaces the plane wave dispersion
relation in a homogeneous medium. Suffice to add that with the aid
of (\ref{InhomStatDiII})-(\ref{InhomStatVacIII}), one may now
explicitly construct the Maxwell and excitation 2-forms in all
three regions, and hence the electromagnetic field 1-forms
$\{\Me{U}, \Mb{U}, \Md{U}, \Mh{U}\}$ in all three regions. The
schematic behaviour of the real electric electric field amplitude
in all regions is shown in figure \ref{electricamp}. Solutions
with different incident polarizations follow in a similar manner.

Once one has a complete Maxwell solution for a given incident wave
this can be substituted into any $K$-drive in order to calculate
time-averaged force and torque pressures on any area of the slab
in terms of the incident wave parameters $\man{A}_{0},\omega$ and
the medium characteristics $\ep{r}(z), \ell$. For the linear
polarized plane waves (\ref{InhomStatVacI}) and
(\ref{InhomStatVacIII}) one finds non-zero time-averaged
integrated \EM torques in any finite volume of the medium
associated with the Minkowski, symmetrized Minkowski and Abraham
drive forms. However they are all the same. This is to be expected
since, if one observes from the table in Appendix A of paper I
that with the medium at rest ($U=V$), although the medium is
inhomogeneous, the associated current 2-forms $\JUK{}$ for
$U(K)=0$ are the same and $\TA{\dotRU{K}}=0$ for harmonic waves.
Moreover if one considers incident circularly polarized harmonic
plane waves the total time-averaged torques are each zero. Hence
harmonic plane waves cannot distinguish the effects of the
Minkowski, symmetrized Minkowski and Abraham \EM \SEM tensors when
incident on non-accelerating inhomogeneous slabs of simple media.
It is natural to enquire how this result may change when the
medium accelerates. This is explored in the next section for a
simple planar inhomogeneous dielectric slab in
uniform rotation about its axis of symmetry.\\

\section{Circularly-Polarized Plane Waves Incident on a Simple Rotating Inhomogeneous Dielectric Cylinder}
In the last section the fields in a stationary linearly
inhomogeneous slab excited by an incident plane harmonic wave were
found. Such fields yield a non-zero time-averaged net force (and
torque) on such a slab that is the same for both the Abraham and
symmetrized Minkowski \EM \SEM tensors. One could attempt to find
the fields excited by an incident \EM pulse such as that produced
by a laser. Since such fields are not harmonic they may yield
forces and torques that distinguish between such tensors
particularly if one also employs a slab with anisotropic or
magneto-electric properties \cite{GCM8}. 
However the precise nature of the fields in a laser pulse with a
finite spot size is clearly more difficult to ascertain
theoretically and this makes the problem of matching fields at the
slab interfaces more difficult than with harmonic plane waves.

%
In this section it is shown how circularly polarized {\it plane
harmonic} waves can excite a transverse plane wave field in a
simple inhomogeneous {\it uniformly rotating} medium analogous to
that found in the previous section. 
The form of this wave is determined by a single amplitude function
of $z$ in the cylindrical coordinates used in \S\ref{TBDsect},
satisfying a second order {\it ordinary} differential equation
with coefficients dependent on the frequency of the incident wave,
the speed of rotation of the slab and its constitutive properties.
Although this equation has no solution in terms of simple analytic
functions it is amenable to numerical analysis. Furthermore one
can match a basis of such solutions to the vacuum plane waves at
the rotating plane slab interfaces. If one accepts that the edge
effects produced by the rim of the rotating slab are ignorable
then an approximate numerical estimate of the fields excited in
the slab by normally incident harmonic plane waves can be made.
From such a solution one may compute the net time-averaged torque
on the slab for both the Abraham and symmetrized Minkowski \EM
\SEM tensors. It will be shown that for a simple medium with
relative permeability one and a permittivity that varies linearly
with $z$ in the slab these two torques are significantly
different.


In cylindrical coordinates, suppose the cylindrical slab rotates
with constant angular speed $\Omega$ about its axis of symmetry
along the $z$-direction. Let the cylinder domain $II$ have
constant permeability $\mu_{r}$ but relative permittivity
$\ep{r}(z)$ and denote the left and right vacuum domains by $I$
and $III$ respectively. Since the field in this medium is to be
excited by a transverse circularly polarized plane harmonic vacuum
wave we look for monochromatic time-harmonic interior solutions
with a TDB-type field ans\"{a}tz in the cylindrical coframe used
in \S\ref{TBDsect}:
\begin{eqnarray*}
    \Mb{U}_{II} &=& {\cc}^{-1}\,\man{E}B(z)\Exp{ \ri\theta - \ri\omega t }(e^{1} + \ri e^{2})  \\
    \Md{U}_{II} &=& \ep{0}\man{E}D(z)\Exp{ \ri\theta - \ri\omega t }(e^{1} + \ri e^{2}).
\end{eqnarray*}
With the real constant $\man{E}$ assigned  the physical dimensions
of $\frac{[Q]}{[L^2][\ep{0}]}$, the complex amplitudes $B(z),D(z)$
to be determined are dimensionless. The electric field 1-form
$\Me{U}_{II}$ and the magnetic field 1-form $\Mh{U}_{II}$ are now
determined from the constitutive relation for a rotating
dielectric. Generalizing the $m=0$, TDB mode in homogeneous media
found in \S\ref{TBDsect} they may be written in the form.
\begin{eqnarray*}
    \Me{U}_{II} &=& \man{E}\left(E_{1}(r,z)e^{1} + E_{2}(r,z)e^{2} + E_{3}(r,z)e^{3}\right) \, \Exp{\ri\theta - \ri\omega\,t }   \\
    \Mh{U}_{II} &=& \cc\ep{0}\man{E}\,\left(H_{1}(r,z)e^{1} + H_{2}(r,z)e^{2} + H_{3}(r,z)e^{3}\right)\, \Exp{\ri\theta - \ri\omega\,t } .
\end{eqnarray*}
Using the constitutive relations (2.23) of paper I with $\V=\Omega
\PD{\theta}$ it follows that:
\begin{eqnarray*}
    E_{1}(r,z) = \frac{D(z)}{\ep{r}(z)} \qquad && \qquad H_{1}(r,z) = \frac{B(z)}{\mu_{r}}\\
    E_{2}(r,z) = \ri E_{1}(r,z) \qquad && \qquad H_{2}(r,z) = \ri H_{1}(r,z) \\
    E_{3}(r,z) = -\frac{r B(z)\left(1 -\ep{r}(z)\mu_{r} \right)\Omega}{\cc\mu_{r}\ep{r}(z)}\qquad && \qquad H_{3}(r,z) = -\frac{r D(z)\left(\ep{r}(z)\mu_{r} - 1\right)\Omega}{\cc\mu_{r}\ep{r}(z)},
\end{eqnarray*}
to leading order in $\frac{\nu}{\cc}$. The fields $F^{II}$ and
$G^{II}$ can now be constructed to satisfy $d\, F^{II}=0$ and
$d\,\star G^{II}=0$. These equations require
\begin{eqnarray*}
    \left(\frac{D(z)}{\ep{r}(z)}\right)' &=&  \frac{B(z)}{\cc\mu_{r}\ep{r}(z)}\left( \df \ep{r}(z)\mu_{r}(\Omega-\omega) - \Omega \right)  \\
    B'(z) &=&  -\frac{D(z)}{\cc\ep{r}(z)}\left( \df \ep{r}(z)\mu_{r}(\Omega-\omega) - \Omega \right),
\end{eqnarray*}
where $f'(z)$ denotes the derivative of   $f(z)$ with respect to
$z$. Substituting $D(z)$  into the first differential equation
gives a second order differential equation for $B(z)$:
\begin{eqnarray}\label{masterODE}
    B''(z) -\frac{\mu_{r}(\Omega - \omega)\ep{r}'(z)B'(z)}{\ep{r}(z)\mu_{r}(\Omega-\omega) - \Omega } + \frac{\left( \df \ep{r}(z)\mu_{r}(\Omega-\omega) - \Omega \right)^{2}B(z)}{\cc^{2}\mu_{r}\ep{r}(z)}  =0 .
\end{eqnarray}
Note, that  with a real $\ep{r}(z)$, the coefficients of this differential
equation are real. Write the solution for $B(z)$ as
\begin{eqnarray}\label{Bsigma}
    B(z) &=& C_{1}\sigma_{1}(z) + C_{2}\sigma_{2}(z),
\end{eqnarray}
in terms of  complex dimensionless constants $C_{1},C_{2}$ and a
basis of {\it real} solutions $\{\sigma_{1}(z),\sigma_{2}(z)\}$.
It follows that
\begin{eqnarray*}
    D(z) &=& -\frac{\cc\ep{r}(z)\left(C_{1}\sigma_{1}'(z) + C_{2}\sigma_{2}'(z) \right)}{\df \ep{r}(z)\mu_{r}(\Omega-\omega) - \Omega }.
\end{eqnarray*}
The forms $F^{II}$ and $G^{II}$ inside the rotating slab are now
expressed in terms of a basis of solutions for (\ref{masterODE})
as:
\begin{eqnarray}
\begin{split}\label{FGrot}
    F^{II} =& \frac{\cc\left( C_{1}\sigma_{1}'(z) + C_{2}\sigma_{2}'(z) \right)}{\ep{r}(z)\mu_{r}(\Omega-\omega) - \Omega}\man{E}\Exp{\ri\theta - \ri\omega t }e^{0} \w \left(e^{1} + \ri e^{2}\right) \\
      & + \left(C_{1}\sigma_{1}(z) + C_{2}\sigma_{2}(z)\right)\man{E}\Exp{\ri\theta - \ri\omega t}\left(\ri e^{1} - e^{2}   \right)\w e^{3} \\
      & + \frac{r\Omega\left(1-\ep{r}(z)\mu_{r}\right)\left(C_{1}\sigma_{1}(z) + C_{2}\sigma_{2}(z)\right)}{\cc\ep{r}(z)\mu_{r}}\man{E}\Exp{\ri\theta - \ri\omega t }e^{0} \w e^{3} \\
      & \\
    G^{II} =& \frac{\ep{0}\ep{r}(z)\cc\left( C_{1}\sigma_{1}'(z) + C_{2}\sigma_{2}'(z) \right)}{\ep{r}(z)\mu_{r}(\Omega-\omega) - \Omega}\man{E}\Exp{\ri\theta - \ri\omega t }e^{0} \w \left(e^{1} + \ri e^{2}\right) \\
      & + \frac{\ep{0}}{\mu_{r}} \left(C_{1}\sigma_{1}(z) + C_{2}\sigma_{2}(z)\right)\man{E}\Exp{\ri\theta - \ri\omega t}\left(\ri e^{1} - e^{2}   \right)\w e^{3} \\
      & + \frac{\ep{0} r\Omega\left(1-\ep{r}(z)\mu_{r}\right)\left(C_{1}\sigma_{1}'(z) + C_{2}\sigma_{2}'(z)\right)}{\mu_{r}\left(\df \ep{r}(z)\mu_{r}(\Omega-\omega) - \Omega \right)}\man{E}\Exp{\ri\theta - \ri\omega t }e^{1} \w e^{2}.
\end{split}
\end{eqnarray}
As we have seen in the last section, when $\Omega=0$ and
$\ep{r}(z)$ varies linearly with $z$, the basis solutions can be
expressed in terms of Airy functions. The complex constants in
(\ref{FGrot}) are determined as before by matching $F^{II}$ and
$G^{II}$ to harmonic plane waves at the interfaces of region $II$
with regions $I$ and $III$. Thus with $F^{I}=d\,A^{I}$,
$F^{III}=d\,A^{III}$, $G^{I}=\ep{0}F^{I}$, $G^{III}=\ep{0}
F^{III}$ and
\begin{eqnarray*}
    A^{I} &=& \frac{\cc\man{E}}{\ri\omega}\Exp{\ri\theta - \ri\omega t}\left(\Exp{\ri k^{I}_{R}z}+ E^{I}_{L}\Exp{\ri k^{I}_{L}z}\right)\left(e^{1} + \ri e^{2}\right) \\
          && \\
    A^{III} &=& \frac{\cc\man{E}}{\ri\omega}E^{III}_{R}\Exp{\ri\theta - \ri\omega t + \ri k^{III}_{R}z}\left(e^{1} + \ri e^{2}\right),
\end{eqnarray*}
the plane interface conditions yield the following equations for
the dimensionless complex variables $E^{I}_{L}, C_{1}, C_{2},
E^{III}_{R}$:\\

$\left(\begin{array}{cccc}
     \ds  1
   & \ds -\frac{\sigma_1'(0)}{\lambda(0)} 
   & \ds -\frac{\sigma_2'(0)}{\lambda(0)} 
   & \ds 0           
\\
   \ds \ri\mu_r     
 & \ds \sigma_1(0) 
 & \ds \sigma_2(0) 
 & \ds 0
 \\
   \ds 0 
 & \ds -\frac{\sigma_1'(\ell)}{\lambda(\ell)}  
 & \ds  -\frac{\sigma_1'(\ell)}{\lambda(\ell)} 
 & \ds  \Exp{\frac{\ri\omega\ell}{\cc}} 
\\
   \ds 0
 & \ds \sigma_1(\ell) 
 & \ds \sigma_2(\ell) 
 & \ds -\ri\mu_r\Exp{\frac{\ri\omega\ell}{\cc}} 
\end{array}\right)
\left(\begin{array}{c}
   E^{I}_{L}\\
   C_{1}\\
   C_{2} \\
   E^{III}_{R}
\end{array}\right) =
\left(\begin{array}{c}
    -1\\        
   \ds \ri\mu_r \\
   \ds 0\\        
   0
\end{array}\right) \label{CHANGE1}$ \\ \quad \\
where
\begin{eqnarray*}
    \lambda(z) &=& \frac{\ep{r}(z)\mu_{r}\left(\Omega - \omega\right) - \Omega}{\cc} .
\end{eqnarray*}
This matrix equation admits a  solution provided
\begin{eqnarray*}
   \Exp{\ri\frac{\omega}{\cc}\ell}\,\bigg(  \left( \sigma_{1}(\ell)\sigma_{2}(0)  - \sigma_{1}(0)\sigma_{2}(\ell) \df\right) + \frac{\mu_{r}^{2}\left( \df \sigma_{1}'(\ell)\sigma_{2}'(0) - \sigma_{1}'(0)\sigma_{2}'(\ell) \df\right)}{\lambda(0)\lambda(\ell)}  \label{CHANGE2}
\\
   -\, \ri\mu_r\, \frac{\left( \df \sigma_{1}'(\ell)\sigma_{2}(0)  -
       \sigma_{1}(0)\sigma_{2}'(\ell)\right)}{\lambda(\ell)} -
     \ri\mu_r\,\frac{\left(  \df \sigma_{1}'(0)\sigma_{2}(\ell)  - \sigma_{1}(\ell)\sigma_{2}'(0) \right)}{\lambda(0)}  \bigg) \,\neq 0 \label{CHANGE3}.
\end{eqnarray*}
Thus one has a matched solution to the complete  \EM system in
terms of a basis of solutions for (\ref{masterODE}) and
$\ep{r}(z)$. With $K=\PD{\theta}$, one may now express the
time-averaged total torques, $-\int_{\calV}\, \TA{ i_{U}\,d\tauK{AB} }$
and $-\int_{\calV}\, \TA{ i_{U}\,d\tauK{SM} }$, on a thin rotating disk
of volume $\calV$ in terms of this solution and for each $\Omega$
and $\omega$ numerically integrate (\ref{masterODE}) to find the
needed values of $\sigma_{1}(z)$ and $\sigma_{2}(z)$ in the slab.

The results in figures \ref{AbrPlotHigh} and \ref{SMPlotHigh} are
for $\ep{r}(z)= \alpha + \frac{\beta}{\ell}\,z$ with real
constants $\alpha,\beta$ describing the variation of permittivity
across the cylindrical disc of thickness $\ell$. They clearly
indicate that if one could devise an experiment to measure the
mechanical torque needed to maintain a simple inhomogeneous
dielectric medium in uniform rotation both in the absence and
presence of a normally incident circularly polarized harmonic
plane wave then one may be able to discriminate between the
effects induced by different \EM \SEM tensors.

\section{Conclusions}
Unlike the Minkowski tensor and its symmetrized version, the
divergence of the Abraham tensor depends {\it explicitly} on the
bulk acceleration field of the medium. Given simple \EM
constitutive relations this implies the existence of local
acceleration dependent Abraham forces and torques in the presence
of \EM fields. One could write the Abraham tensor as the sum of
any other symmetric tensor (including the symmetrized Minkowski
tensor) and the difference between the two. This of course makes
no difference to the divergence but {\it assigning} the difference
in the tensors to the accompanying matter tensor appears to us ad
hoc and unwarranted. As an illustration we would prefer to
identify the radiation reaction force experienced by  a radiating
point charge (that depends on the rate of change of the
acceleration of the particle) with an intrinsic property of the
charge's interaction with the \EM field rather than with the
charge's inertia (that conventionally depends only on the
acceleration) and some additional force. Furthermore from an
action variational viewpoint it is natural to associate particular
\SEM tensors with particular interacting systems since the extrema
of the total action determine both the total \SEM {\it and} the
field equations for the fully coupled system.

The view advocated here is that it does make sense to try and pin
down by theoretical modelling and  experiment the most appropriate
{\it form} of an  \EM \SEM tensor for as large a range of \EM
constitutive relations as possible. As emphasized in
\cite{DGT,DGT2}, 
we believe that variational principles including gravitation are a
valuable theoretical guide in this endeavor. On the experimental
side it is suggested here that experiments involving accelerating
media might offer a sensitive means to discriminate between
alternatives.

\begin{figure}[p]
    \centering
    \includegraphics[width=14cm]{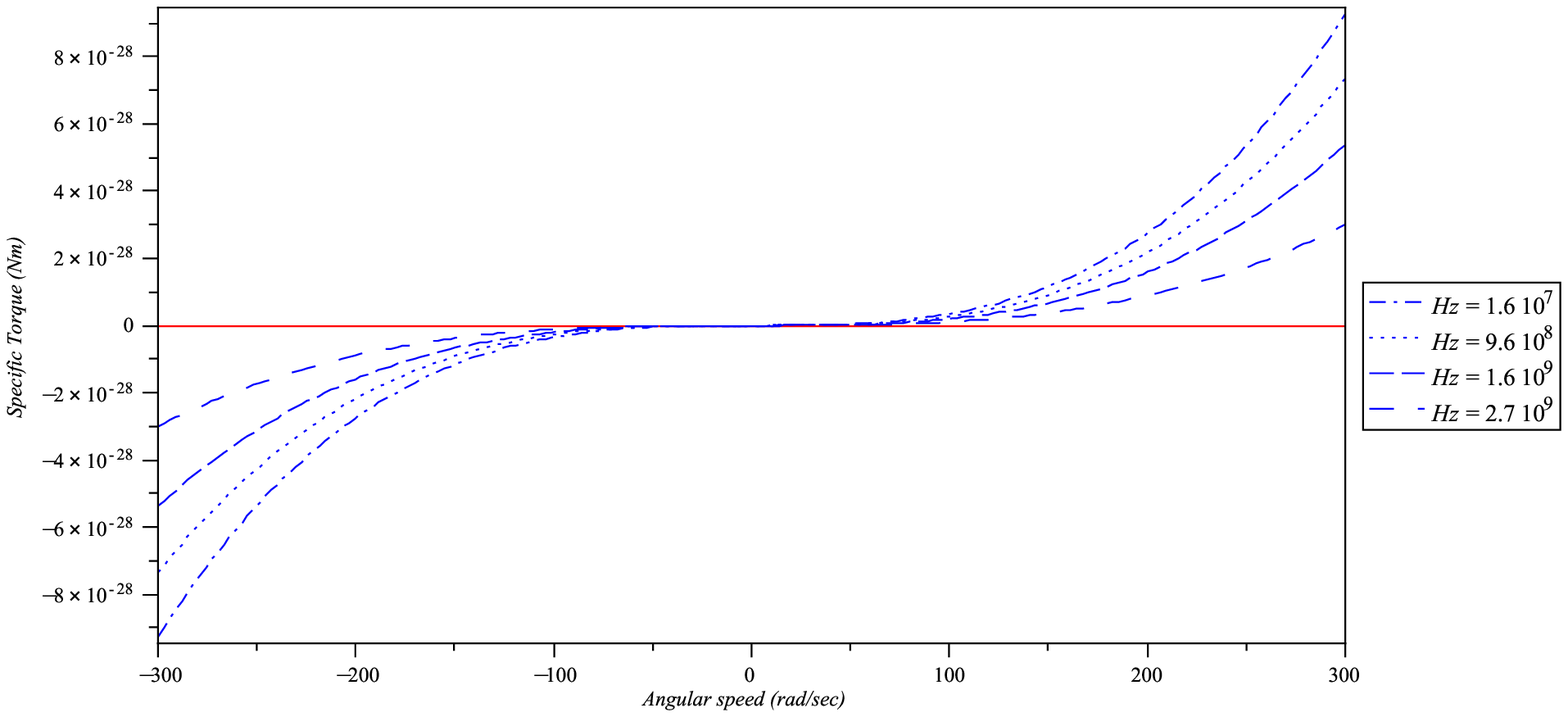}
    \caption{Variation of  time-averaged specific torques  with angular speed of a thin rotating cylinder calculated with the {\bf Abraham} \EM \SEM tensor. The physical torque is obtained by multiplying  each specific torque by  ${\cal E}^2$ where ${\cal E}$ is the electric field amplitude in $volts\, m^{-1}$ of the normally incident plane circularly polarized harmonic wave with frequency (including the optical range) given by the legend on the right of the figure.}
    \label{AbrPlotHigh}
\end{figure}

\begin{figure}[p]
    \centering
    \includegraphics[width=14cm]{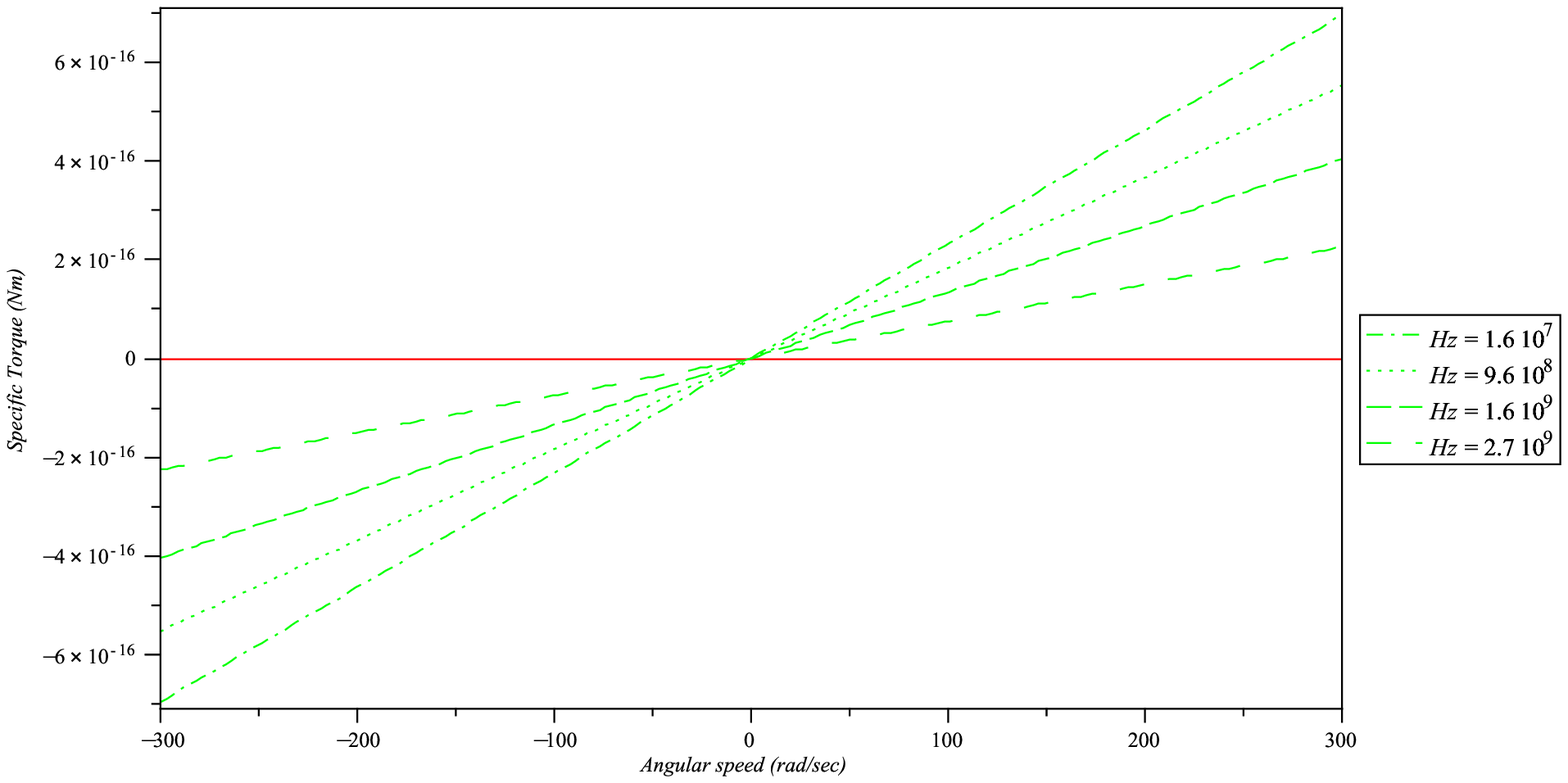}
    \caption{Variation of  time-averaged specific torques  with angular speed of a thin rotating cylinder calculated with the {\bf symmetrized Minkowski} \EM \SEM tensor. The physical torque is obtained by multiplying  each specific torque by  ${\cal E}^2$ where ${\cal E}$ is the electric field amplitude in $volts\, m^{-1}$  of the normally incident plane circularly polarized harmonic wave with frequency (including the optical range) given by the legend on the right of the figure.}
    \label{SMPlotHigh}
\end{figure}

From the calculations for the macroscopic time-averaged specific
torques on simple media shown in figures \ref{AbrPlotHigh} and
\ref{SMPlotHigh}, it is clear that those corresponding to the
symmetrized Minkowski tensor are much larger than those associated
with the Abraham tensor. In each figure the physical torque has
been calculated for a cylinder of radius  $1\, m$ and thickness
$\ell=1 \,mm$ as a function of the uniform rotation speed $\Omega$
in radians per second. The parameters defining the linear
variation of permittivity  across the thickness of the cylinder
are $\alpha = 1$ and $\beta = 100$. The torque curves are
calculated for a range of frequencies (indicated in $Hz$ in the
legend on the right of each figure) of the normally incident
circularly polarized incident plane wave. The magnitude of each
physical torque (in $Newton\, m$) is obtained by multiplying the
indicated specific torques by $\man{E}^{2}$ where $\man{E}$ is the
amplitude of the electric field of the incident plane wave in
$volts\, m^{-1}$. Recall that the time-averaged power in a
harmonic plane wave crossing unit normal area is $ \man{E}^{2}/{(2
Z_{0})}$ $watts\, m^{-2}$ in terms of the characteristic impedance
$Z_{0}=\sqrt{{\mu_{0}}/{\ep{0}}}$ of free space. The detection of
such a torque would require the measurement of an externally
applied torque to maintain a non-deformed cylinder in uniform
rotation. It is interesting to note from the figures the torque
asymmetry about $\Omega=0$. This asymmetry can be shifted along
the horizontal axes by using an oppositely polarized incident beam
and is clearly a direct manifestation of its intrinsic angular
momentum. The significant quantitative differences between the
time-averaged Abraham and symmetrized Minkowski torques  in simple
inhomogeneous rotating dielectrics  suggest that new opportunities
exist for exploring the effects of classical electrodynamics in
more general rotating media.\\

\section{Acknowledgements}
The authors are grateful to the Cockcroft Institute, the Alpha-X
project, STFC and EPSRC (EP/E001831/1) for financial support for
this research.



\appendix{}
\section{Notation}
The natural mathematical language to discuss the differential
properties of tensor fields on spacetime and their relation to
integrals over material domains is in terms of differential forms
and their associated exterior calculus
(Benn \& Tucker 1988). In this section a brief summary is given of
the relevant notation used in subsequent sections. A key concept
throughout involves the role of the spacetime metric tensor field
and possible isometries that it may possess. A spacetime metric
tensor field $\g$ is a symmetric bilinear form on spacetime that
can always be represented in a local cobasis of differential
1-forms $\{e^{a}\}$ as
\begin{eqnarray}
    \g &=& -e^{0}\tensor e^{0} +  e^{1}\tensor e^{1} + e^{2}\tensor e^{2} + e^{3}\tensor e^{3}.
\end{eqnarray}
If $\{X_{b}\}$ is the dual local basis of vector fields on
spacetime defined so that $e^{a}(X_{b})=\delta^{a}_{b}$
$(a,b,=0,1,2,3)$ one has the induced contravariant metric
\begin{eqnarray}
    \ginv &=& -X_{0}\tensor X_{0} +  X_{1}\tensor X_{1} + X_{2}\tensor X_{2} + X_{3}\tensor X_{3}.
\end{eqnarray}
If $\alpha$ is any given 1-form, one has an associated vector
field $\wt{\alpha}$ defined so that $\beta(\wt{\alpha})=
\ginv(\alpha,\beta)$ for all 1-forms $\beta$.  Since $\ginv$ is
symmetric this will be abbreviated $\wt{\alpha}=\ginv(\alpha)$. In
a similar way if $X$ is any given vector field one has an
associated 1-form $\wt{X}= \g(X)$.

In the $\g$-orthonormal basis $\{e^{a}\}$ one has a canonical
local 4-form denoted $\star 1$ and defined to be $e^{0}\w e^{1} \w
e^{2} \w e^{3}$. The Hodge map $\star$ induced by $\star 1$ maps
$p$-forms to $(4-p)$-forms on spacetime
(Benn \& Tucker 1988; see also Appendix A of paper I). The metric
also uniquely defines the covariant derivative $\nabla_{X}$ with
respect to any vector field $X$. This has the property
$\nabla_{X}\g=0$ and $\nabla_{X}Y - \nabla_{Y}X = [X,Y]$ for all
vector fields $X,Y$. In this expression $[X,Y]$ denotes the
commutator bracket. While $\nabla_{X}$ has a type-preserving
action on any tensor field the exterior derivative $d$ is defined
to act only on antisymmetric tensor fields (differential forms)
and has the property $d \circ d=0$. Contraction of any $p$-form
$\beta$ with $X$ is denoted $i_{X}\beta$. The interior operator
$i_{X}$ is a graded derivation defined so that
\begin{eqnarray}
    i_{X}(\alpha \w \beta) &=& (i_{X}\alpha) \w \beta + (-1)^{p}\, \alpha \w i_{X}\beta,
\end{eqnarray}
for any $p$-form $\alpha$ and $q$-form $\beta$. If $p=1$, one
defines $i_{X}\alpha=\alpha(X)$, and if $p=0$, $i_{X}\alpha=0$,
for all vector fields $X$. One has the useful relations
\begin{eqnarray*}
    \star\star\Phi &=& (-1)^{p+1}\Phi \\
    i_{X}\star\Phi &=& \star ( \Phi \w \wt{X} ) \\
    \Lie_{X}\Phi &=& i_{X}d\Phi + di_{X}\Phi,
\end{eqnarray*}
for any $p$-form $\Phi$ on spacetime where $\Lie_{X}$ denotes Lie
differentiation with respect to $X$.

The spacetime divergence operator $\nabla\cdot$ takes a simple
form if one uses the $\g$-orthonormal basis above. Acting on a
symmetric covariant tensor $T$ it defines by contraction  on the
first argument the $1$-form
\begin{eqnarray} \label{DIVDEF}
    \nabla\cdot T &=& \sum_{a=0}^{3}(\nabla_{X_{a}}\, T) ( X^{a},-)
\end{eqnarray}
where $\{X^{a}\}=\{-X_{0},X_{1},X_{2},X_{3}\}$. For a symmetric
tensor $T$ it is sufficient to write the right hand side as
$(\nabla_{X_{a}}\, T) ( X^{a})$. A local spacetime isometry with
respect to $\g$ is a local diffeomorphism that preserves this
metric. A vector field $K$ that generates such a diffeomorphism is
called a Killing vector field and it satisfies $\Lie_{K}\g=0$ in
terms of the operation of Lie differentiation with respect to $K$.
The operators $\star, d, i_{X}, \nabla_{X}, \Lie_{X}$ offer a
powerful computational tool-kit when working with differential
forms.

For any {\it smooth} $p$-form $\Phi$ in a bounded regular region
$\M$ of a manifold one can express the integral of $d\Phi$ over
$\M$ in terms of the integral of $\Phi$ over the boundary
$\PD{}\M$ of $\M$:
\begin{eqnarray}
    \int_{\M}\, d\Phi &=& \int_{\PD{}\M}\, \Phi .
\end{eqnarray}
This is a statement of Stokes' theorem for $p$-forms.

The Gibbs calculus of vector fields in 3-dimensional Euclidean
space is readily exposed by correspondences induced by the
exterior operations above. A space of 3-dimensions may be
considered as a particular hypersurface in spacetime. If the
metric above induces the metric
\begin{eqnarray}
    {\mathbf \g} &=& e^{1}\tensor e^{1} + e^{2}\tensor e^{2} + e^{3}\tensor e^{3}
\end{eqnarray}
on this hypersurface it is Euclidean and one may introduce the
Euclidean canonical form $\# 1= e^{1}\w e^{2} \w e^{3}$ by
restriction. Since spacetime is assumed  time-oriented one may
employ a future-pointing timelike unit vector field $U$ on
spacetime (with $\g(U,U)=-1$ and $\wt{U}=e^{0}$) to fix a coherent
orientation by relating $\#1$ to $\star 1$ by
\begin{eqnarray}
    \star 1 &=& \wt{U} \w \#1 .
\end{eqnarray}
If the hypersurface is given as $t=$ constant for some time
coordinate $t$, an inertial frame exists in Minkowski spacetime
with $U=\frac{1}{\cc}\pdiff{}{t}=-X_{0}$ such that $\nabla\,U=0$.
Throughout this article the constant $\cc$ denotes the speed of
light in the vacuum. Any $p$-form $\alpha$ on spacetime is termed
spatial with respect to such a $U$ if $i_{U}\alpha=0$. An over-dot
will denote (Lie) differentiation with respect to the coordinate
$t$  so
\begin{eqnarray}
    \dot{\alpha} &\equiv& \cc\Lie_{U}\alpha = \Lie_{\PD{t}}\alpha.
\end{eqnarray}
On a Euclidean hypersurface in spacetime, exterior differentiation
of spatial $p$-forms $\phi$ is denoted  $\sd\phi$ such that
\begin{eqnarray}\label{LIE}
    d\phi &=& \sd\phi - \wt{U} \w \Lie_{U}\phi = \sd\phi + dt \w \Lie_{\PD{t}}\phi.
\end{eqnarray}
One has the following relations
\begin{eqnarray*}
    \#1 &=& -i_{U}\star 1 = -\star \wt{U} \\
    \# \# \phi &=& \phi
\end{eqnarray*}
for all spatial $p$-forms $\phi$. Furthermore if $\Evec, \Ewec$
denote Euclidean vector fields in the Gibbs notation corresponding
to the vector fields $\wt{v}, \wt{w}$ for some spatial 1-forms $v,
w$ then
\begin{eqnarray}
    \label{EDIV} \text{div } \Evec \qquad &&\text{corresponds to} \qquad \wt{\# \sd \# v}\; ; \\
\nonumber && \\
    \label{CURL} \text{curl } \Evec \qquad &&\text{corresponds to} \qquad \widetilde{\# \sd v}\; ; \\
\nonumber && \\
    \label{VPRODUCT} {\Evec \times \Ewec} \qquad &&\text{corresponds to} \qquad \widetilde{ \# ( v \w w )}\; ;\\
\nonumber && \\
    \label{GRAD} \text{grad } \psi \qquad &&\text{corresponds to} \qquad \wt{\sd \psi},
\end{eqnarray}
for any scalar $\psi$  field on spacetime.\\

\section{Electromagnetic Fields on Spacetime}
We suppose that a polarizable material continuum is given in terms
of a set of piecewise smooth material properties that determine
its  interaction with classical gravitational and \EM fields. The
classical macroscopic Maxwell system for the electromagnetic
2-form $F$ on spacetime can be written as \cite{GCM8}:
\begin{eqnarray}\label{MAXSYS}
    dF = 0 \qquadand d\star G = j,
\end{eqnarray}
where the excitation 2-form $G$ depends on the interaction with
the medium  and the 3-form electric 4-current $j$ encodes the
electric charge and current source\footnote{
    All electromagnetic tensors in this article have
    dimensions constructed from the SI dimensions $[M], [L], [T], [Q]$ where $[Q]$
    has the unit of the Coulomb in this system. We adopt $[\g]=[L^{2}],
    [G]=[j]=[Q],\,[F]=\frac{[Q]}{[\ep{0}]}$ where the permittivity of free space
    $\ep{0}$ has the dimensions $[ Q^{2} T^{2} M^{-1} L^{-3}]$ and
    $\cc=\frac{1}{\sqrt{\ep{0}\mu_{0}}}$ denotes the speed of light in vacuo.
    Note that the operators $d$ and $\nabla$  preserve the physical dimensions of tensor fields but with $[\g ]=[L^{2}]$, for $p$-forms $\alpha$ in $4$ dimensions,
    one has $[\star \alpha]=[\alpha] [L^{4-2p}]$.}.
Such an electric 4-current describes both (mobile) electric charge
and effective (Ohmic) currents in a conducting medium. To close
this system in a background gravitational field, {\it
electromagnetic constitutive relations} relating $G$ and $j$ to
$F$ are necessary.

The history of a particular observer field in spacetime is
associated with an arbitrary {\it unit} future-pointing timelike
4-velocity vector field $U$. The field $U$ may be used to describe
an {\it observer frame} on spacetime and its integral curves model
{\it idealized observers}. An orthogonal decomposition of $F$ with
respect to any observer field $U$ gives rise to a pair of {\it
spatial} 1-forms on spacetime. The 1-form spatial {\it electric
field} $\Me{U}$ and 1-form spatial {\it magnetic induction field}
$\Mb{U}$ associated with $F$ are defined with respect to an
observer field $U$ by
\begin{eqnarray}\label{intro_e_b}
    \Me{U} = i_{U}F \qquadand \cc\Mb{U} = i_{U} \star F.
\end{eqnarray}
Since $\g(U,U)=-1$ and $i_{U}\Me{U}=i_{U}\Mb{U}=0$:
\begin{eqnarray}\label{intro_F}
    F &=& \Me{U}\w\wt{U} - \star\,(\cc\Mb{U}\w \wt{U}).
\end{eqnarray}
Likewise the 1-form spatial {\it displacement field} $\Md{U}$ and
the 1-form spatial {\it magnetic field} $\Mh{U}$ associated with
$G$ are defined with respect to $U$ by
\begin{eqnarray}\label{Media_d_h}
    \Md{U} = i_{U}G  \qquadand \frac{\Mh{U}}{\cc} = i_{U}\star G,
\end{eqnarray}
so
\begin{eqnarray}\label{Media_G}
    G &=& \Md{U}\w \wt{U} - \star\left( \frac{\Mh{U}}{\cc}\w \wt{U} \right),
\end{eqnarray}
with $i_{U}\Md{U}=i_{U}\Mh{U}=0$. At the history of any sharp
interface between different media, given as the piecewise smooth
(non-null) spacetime hypersurface $f=0$, the system of Maxwell
equations is supplemented by interface conditions on the fields
$F$ and $G$
\begin{eqnarray}\label{MaxwellBC}
\begin{split}
    \left.\df[F]\right|_{f=0} \w df &= 0 \\
    \left.\df[\star G]\right|_{f=0} \w df &= j_{s},
\end{split}
\end{eqnarray}
where $[H]$ denotes the discontinuity in the field $H$ across the
hypersurface. The 3-form $j_{s}$ on the hypersurface is non zero
if it supports a real current 3-form there.\\

\section{Time-Dependent Maxwell Systems in 3-Space}
The spacetime description above is natural for the Maxwell system
since it makes no reference to any particular frame in spacetime.
However to make contact with descriptions in particular frames or
non-relativistic formulations a reduction in terms of frame
dependent fields becomes mandatory. The spacetime Maxwell system
can now be reduced to a family of parameterized exterior systems
on $\real^{3}$. Each member is an exterior system involving forms
on $\real^{3}$ depending parametrically on some time coordinate
$t$ associated with $U$. Let the $(3+1)$ split of the electric
4-current 3-form with respect to a foliation  of spacetime by
spacelike hypersurfaces with constant $t$  be
\begin{eqnarray}\label{decompJ}
    j &=& \frac{\J{}}{\cc} \w \wt{U} + \RU{},
\end{eqnarray}
with $i_{U}\J{}=i_{U}\RU{}=0$ and $\RU{}=\hatRU{}\# 1$, where
$\J{},\RU{}$ are the spatial electric current density 2-form and
spatial electric charge density 3-form respectively. The
differential operator $\sd$ on spacetime forms is adapted to those
spacetimes (such as Minkowski spacetime where gravity is absent)
that can be foliated by hypersurfaces with constant coordinate $t$
where $U=\frac{1}{\cc}\PD{t}$. Then, from (\ref{MAXSYS}),
\begin{eqnarray}\label{dj}
    dj &=& 0
\end{eqnarray}
yields
\begin{eqnarray}\label{cont}
    \sd\man{J}^{U} + \dot{\rho}^{U} &=& 0.
\end{eqnarray}
 The $(3+1)$ split of the
spacetime covariant Maxwell equations (\ref{MAXSYS}) with respect
to $\wt{U}=-\cc dt$ becomes
\begin{eqnarray}
     \label{M1} \sd\,\Me{U} &=& -\dotMB{U} \\
     \label{M2} \sd\,\MB{U} &=& 0 \\
     \label{M3} \sd\,\Mh{U} &=& \J{} + \dotMD{U} \\
     \label{M4} \sd\,\MD{U} &=& \RU{}
\end{eqnarray}
where $\MD{U}=\#\Md{U}$ and $\MB{U}=\#\Mb{U}$. All $p$-forms
($p>0$) in these equations are independent of $dt$, but have
components that may depend parametrically on $t$.\\

\section{Electromagnetic Constitutive Relations}
The two 2-forms $F$ and $G$ in the macroscopic Maxwell equations
on spacetime are fundamentally related by smoothing the
microscopic sources of  the \EM  fields in the medium. In many
circumstances one then relies on phenomenological relations for
closure relations. In such relations the excitation form $G$ is in
general a functional (possibly non-local in spacetime) of the
Maxwell form $F$, its covariant differentials, thermodynamic
properties, deformation properties and the state of motion of the
medium:
\begin{eqnarray*}
    G &=& \man{Z}[F, \nabla\,F\ldots].
\end{eqnarray*}
Such a functional may induce  non-linear and non-local relations
between $\Md{U}, \Mh{U}$ and $\Me{U}, \Mb{U}$. Electrostriction
and magnetostriction arise from the dependence of $\man{Z}$ on the
deformation tensor of the medium and its covariant derivatives.
For general {\it linear continua},  a knowledge of a collection of
{\it constitutive tensor fields} $Z^{(r)}$ on spacetime may
suffice so that
\begin{eqnarray*}
    G &=& \sum_{r=0}^{N}Z^{(r)}[\nabla^{r}F,\ldots].
\end{eqnarray*}
In idealized (non-dispersive) {\it simple continua}, one adopts
the idealized {\it local} relation
\begin{eqnarray*}
    G &=& Z(F),
\end{eqnarray*}
for some degree 4 constitutive tensor field $Z$, parameterized by
scalars that depend on the medium. In the vacuum $G=\ep{0}F$ where
$\ep{0}$ is the constant permittivity of the vacuum. Regular {\it
lossless, non-conducting, linear isotropic media} can be
described by a bulk 4-velocity field $V$ of the medium, a real
relative permittivity scalar field $\ep{r} > 0$ and a real
relative permeability scalar field $\mu_{r} > 0 $. In this case,
the structure of the tensor $Z$ follows from
\begin{eqnarray}\label{DielecCR}
\begin{split}
    G &\eqq \ep{0}\ep{r}i_{V}F \w \wt{V} - \frac{\ep{0}}{\mu_{r}}\star\left(i_{V}\star F \w \wt{V}\right) \\
      &\eqq \ep{0}\left( \ep{r} - \frac{1}{\mu_{r}} \right)i_{V}F \w \wt{V} + \frac{\ep{0}}{\mu_{r}} F.
\end{split}
\end{eqnarray}
For inhomogeneous media the relative permittivity and permeability
scalars $\ep{r}$ and $\mu_{r}$ will not be constants. In a general
frame $U$ comoving with the medium ($U=V$), (\ref{DielecCR})
yields
\begin{eqnarray}\label{dielecComCR}
    \Md{V} = \ep{0}\ep{r}\Me{V} \qquadand \Mh{V} = (\mu_{0}\mu_{r})^{-1}\Mb{V},
\end{eqnarray}
which are the familiar closure relations for simple (idealized)
electrically neutral isotropic non-dispersive polarizable media.\\

\subsection{Minkowski Constitutive Relations for Moving Media}
The above algebraic constitutive relation (\ref{DielecCR})
involves the bulk 4-velocity field of a simple medium. It is
straightforward to find the induced relations between the fields
$\{\Me{U},\Mb{U},\Md{U},\Mh{U}\}$  relative to an observer in a
frame $U \neq V$. From (\ref{intro_F}) and (\ref{Media_G}), one
has
\begin{eqnarray*}
    \Me{V} &=& i_{V}F = \Me{U}(V)\wt{U} - \wt{U}(V)\Me{U} - \star\left( \cc\Mb{U} \w \wt{U} \w \wt{V}\right) \\
    \cc\Mb{V}&=& i_{V}\star F = \star(\Me{U} \w \wt{U} \w \wt{V}) +  \cc\Mb{U}(V)\wt{U} - \cc\wt{U}(V)\Mb{U} \\
    \Md{V} &=& i_{V}G = \Md{U}(V)\wt{U} - \wt{U}(V)\Md{U} - \star\left( \frac{\Mh{U}}{\cc} \w \wt{U} \w \wt{V}\right) \\
    \frac{\Mh{V}}{\cc} &=& i_{V}\star G = \star\left(\Md{U} \w \wt{U} \w \wt{V}\right) + \frac{\Mh{U}(V)}{\cc}\wt{U} - \frac{\wt{U}(V)}{\cc}\Mh{U}.
\end{eqnarray*}
Inserting these in the constitutive relations (\ref{dielecComCR})
yields
\begin{eqnarray*}
\begin{split}
    \mbox{\small $\Md{U}(V)\wt{U} - \wt{U}(V)\Md{U} - \star\left( \frac{\Mh{U}}{\cc} \w \wt{U} \w \wt{V} \right)$} &= \mbox{\small $\ep{}\left[\df \Me{U}(V)\wt{U} - \wt{U}(V)\Me{U} - \star\left( \cc\Mb{U} \w \wt{U} \w \wt{V}\right)\right]$} \\
    \mbox{\small $\cc\star\left(\Md{U} \w \wt{U} \w \wt{V}\right) + \Mh{U}(V)\wt{U} - \wt{U}(V)\Mh{U}$} &= \mbox{\small $\mu^{-1}\!\left[\!\df\star\left(\frac{\Me{U}}{\cc} \w \wt{U} \w \wt{V}\right) +  \Mb{U}(V)\wt{U} - \wt{U}(V)\Mb{U} \right]$},
\end{split}
\end{eqnarray*}
where $\ep{}=\ep{r}\ep{0},\mu=\mu_{r}\mu_{0}$ in terms of the
relative permittivity scalar $\ep{r}$ and relative permeability
$\mu_{r}$. Contracting with $U$ gives the relations
\begin{eqnarray}
\begin{split}\label{UconCR}
    \Md{U}(V) &= \ep{}\Me{U}(V) \\
    \Mh{U}(V) &= \mu^{-1}\Mb{U}(V),
\end{split}
\end{eqnarray}
which yields
\begin{eqnarray}
\begin{split}\label{transCR2}
    \wt{U}(V)\Md{U} + \star\left( \frac{\Mh{U}}{\cc} \w \wt{U} \w \wt{V} \right) &= \ep{}\left[\df\wt{U}(V)\Me{U} + \star\left( c\Mb{U} \w \wt{U} \w \wt{V}\right)\right] \\
    \wt{U}(V)\Mb{U} - \star\left(\frac{\Me{U}}{\cc} \w \wt{U} \w \wt{V}\right)  &= \mu\left[\df\wt{U}(V)\Mh{U} - \cc\star\left(\Md{U} \w \wt{U} \w \wt{V}\right)\right].
\end{split}
\end{eqnarray}
A laboratory Minkowski {\it inertial frame}  is described by
$U=\frac{1}{\cc}\PD{t}$  in  inertial Cartesian coordinates
$\{t,x^1,x^2,x^3\}  $  with $\{e^0=-\cc dt, e^1=d x^1,e^2= d
x^2,e^3= d x^3\} $   and, relative to $U$, the medium 4-velocity
$V$ has the orthogonal decomposition:
\begin{eqnarray*}
    V &=& \gamma\left( U + \frac{\V}{\cc} \right),
\end{eqnarray*}
where the spatial field  $\V$ is the Newtonian velocity
field\footnote{In inertial Cartesian coordinates $\V =
\sum_{j=1}^{3} v^{j}(t,x^{1},x^{2},x^{3}) \pdiff{}{x^{j}}$.} of
the medium relative to $U$,  $\gamma^{-1} \equiv \sqrt{(1-
\frac{\nu^2}{\cc^2})} $ and $\nu^{2}\equiv\g(\V,\V)$. With these
definitions it follows that:
\begin{eqnarray*}
    \g(U,V) = -\gamma \qquad \text{and} \qquad  \wt{U} \w \wt{V} = \frac{\gamma}{\cc}\wt{U} \w  \Vt.
\end{eqnarray*}
Using these in (\ref{transCR2}) and rearranging yields
\begin{eqnarray}
\begin{split}\label{transCR3}
    \Md{U} - \star\left( \wt{U} \w \Vt \w \frac{\Mh{U}}{\cc^{2}}  \right) &= \ep{}\left[\df\Me{U} - \star\left( \wt{U} \w \Vt \w \Mb{U} \right)\right] \\
    \Mb{U} + \star\left(\wt{U} \w \Vt \w  \frac{\Me{U}}{\cc^{2}} \right)  &= \mu\left[\df\Mh{U} + \star\left( \wt{U} \w \Vt \w \Md{U} \right)\right],
\end{split}
\end{eqnarray}
or using the spatial Hodge map  $\#$ defined by $U$:
\begin{eqnarray}
\begin{split}\label{MinkCR}
    \Md{U} + \#\left( \Vt \w \frac{\Mh{U}}{\cc^{2}}  \right) &= \ep{}\left[\df\Me{U} + \#\left( \Vt \w \Mb{U} \right)\right] \\
    \Mb{U} - \#\left( \Vt \w  \frac{\Me{U}}{\cc^{2}} \right)  &= \mu\left[\df\Mh{U} - \#\left( \Vt \w \Md{U} \right)\right].
\end{split}
\end{eqnarray}
These are the constitutive relations first written by Minkowski.
For some purposes it is useful to decouple these expressions and
express $\Me{U}$ and $\Mh{U}$ directly in terms of $\Md{U},\Mb{U}$
and $\V$.\\

\subsection{Minkowski Constitutive Relations for  $\Me{U}(\Md{U},\Mb{U},\V),\Mh{U}(\Md{U},\Mb{U},\V)$}
Taking the exterior product of (\ref{MinkCR}) with $\V$ yields,
\begin{eqnarray*}
    \mu\Vt \w \Mh{U} &=& \Vt \w \Mb{U} - \Vt \w \#\left( \Vt \w  \frac{\Me{U}}{\cc^{2}} \right) + \mu\Vt \w \#\left( \Vt \w \Md{U} \right).
\end{eqnarray*}
For any spatial 1-form  $\alpha^{U}$ with respect to $U$ one has:
\begin{eqnarray}
    \nonumber \Vt \w \#\left( \Vt \w \alpha^{U}\right) &=& -\Vt \w \star\left(\wt{U} \w \Vt \w \alpha^{U} \right) = -\star i_{\V}\left( \wt{U} \w \Vt \w \alpha^{U} \right) \\
    \nonumber                                 &=& \nu^{2}\star \left( \wt{U} \w \alpha^{U} \right) - \alpha^{U}(\V)\star\left( \wt{U} \w \Vt \right) \\
    \label{spatident}                         &=& -\nu^{2}\#\alpha^{U} + \alpha^{U}(\V)\#\Vt.
\end{eqnarray}
Thus
\begin{eqnarray*}
    \mu\Vt \w \Mh{U} &=& \Vt \w \Mb{U} + \frac{\nu^{2}}{\cc^{2}}\#\Me{U} - \frac{\Me{U}(\V)}{\cc^{2}}\#\Vt  - \mu\nu^{2}\#\Md{U} + \mu\Md{U}(\V)\#\Vt \\
                     &=& \Vt \w \Mb{U} + \frac{\nu^{2}}{\cc^{2}}\#\Me{U} - \mu\nu^{2}\#\Md{U} - \frac{1}{\ep{}\cc^{2}}(1-\ep{}\mu \cc^{2})\Md{U}(\V)\#\Vt,
\end{eqnarray*}
using the identity $\Md{U}(\V)=\ep{}\Me{U}(\V)$, obtained by
contracting (\ref{MinkCR}) with $\V$. Since $\#\#=1$
\begin{eqnarray*}
    \mu\#\left(\Vt \w \Mh{U}\right) &=& \#\left(\Vt \w \Mb{U}\right) + \frac{\nu^{2}}{\cc^{2}}\Me{U} - \mu\nu^{2}\Md{U} - \frac{1}{\ep{}\cc^{2}}(1-\ep{}\mu \cc^{2})\Md{U}(\V)\Vt.
\end{eqnarray*}
Substituting this into the first relation of (\ref{MinkCR}) yields
\begin{eqnarray*}
    \left(\ep{}\mu \cc^{2} - \frac{\nu^{2}}{\cc^{2}}\right)\Me{U} &=& \mu \cc^{2}\left(1-\frac{\nu^{2}}{\cc^{2}}\right)\Md{U} - \left(\ep{}\mu \cc^{2} -1\right)\left( \#\left(\Vt \w \Mb{U}\right) - \frac{\Md{U}(\V)}{\ep{}\cc^{2}}\Vt\right).
\end{eqnarray*}
Since
\begin{eqnarray*}
    \frac{1}{\ep{}\mu} &=& \frac{\cc^{2}}{\man{N}^{2}},
\end{eqnarray*}
where $\man{N}^{2}=\ep{r}\mu_{r}$ is the square of the medium
refractive index scalar, this may be written
\begin{eqnarray*}
    \left(\man{N}^{2} - \frac{\nu^{2}}{\cc^{2}}\right)\Me{U} &=& \frac{\man{N}^{2}}{\ep{}}\left(1-\frac{\nu^{2}}{\cc^{2}}\right)\Md{U} - \left(\man{N}^{2}-1\right)\left( \#\left(\Vt \w \Mb{U}\right) - \frac{\Md{U}(\V)}{\ep{}\cc^{2}}\Vt \right).
\end{eqnarray*}
Furthermore from (\ref{MinkCR})
\begin{eqnarray*}
    \ep{}\#\left(\Vt \w \Me{U}\right) &=& \#\left(\Vt \w \Md{U}\right) + \left( \frac{\Mh{U}(\V)}{\cc^{2}}\Vt - \frac{\nu^{2}}{\cc^{2}}\Mh{U} \right) - \ep{}\left(\Mb{U}(\V)\Vt - \nu^{2}\Mb{U}\right) \\
                             &=& \#\left(\Vt \w \Md{U}\right)  - \frac{\nu^{2}}{\cc^{2}}\Mh{U} + \ep{} \nu^{2}\Mb{U} - \frac{1}{\mu \cc^{2}}\left(\ep{}\mu \cc^{2} -1 \right)\Mb{U}(\V)\Vt ,
\end{eqnarray*}
using $\Mb{U}(\V)=\mu\Mh{U}(\V)$, obtained by contracting
(\ref{MinkCR}) with $\V$. Substituting this into the second
relation of (\ref{MinkCR}) yields
\begin{eqnarray*}
    \left(\man{N}^{2} - \frac{\nu^{2}}{\cc^{2}}\right)\Mh{U}  &=& \frac{\man{N}^{2}}{\mu}\left( 1- \frac{\nu^{2}}{\cc^{2}}\right)\Mb{U} - \left(\man{N}^{2}-1 \right)\left(-\#\left( \Vt \w \Md{U} \right) -  \frac{\Mb{U}(\V)}{\mu \cc^{2}}\Vt \right).
\end{eqnarray*}
Thus, the constitutive relations (\ref{DielecCR}) can also be
written
\begin{eqnarray*}
\begin{split}\label{DBeh}
    \left(\man{N}^{2} - \frac{\nu^{2}}{\cc^{2}}\right)\Me{U} & = \frac{\man{N}^{2}}{\ep{}}\left(1-\frac{\nu^{2}}{\cc^{2}}\right)\Md{U} + \left(\man{N}^{2}-1\right)\left( -\#\left(\Vt \w \Mb{U}\right) - \frac{\Md{U}(\V)}{\ep{}\cc^{2}}\Vt \right) \\
    \left(\man{N}^{2} - \frac{\nu^{2}}{\cc^{2}}\right)\Mh{U}  & = \frac{\man{N}^{2}}{\mu}\left( 1- \frac{\nu^{2}}{\cc^{2}}\right)\Mb{U} - \left(\man{N}^{2}-1 \right)\left(-\#\left( \Vt \w \Md{U} \right) -  \frac{\Mb{U}(\V)}{\mu \cc^{2}}\Vt \right).
\end{split}
\end{eqnarray*}
In the non-relativistic limit (to first order in
$\frac{\nu}{\cc}$) these constitutive relations become
\begin{eqnarray}
\begin{split}\label{FOmink}
    \Me{U}  &\;\;\approx\;\;  \frac{\Md{U}}{\ep{0}\ep{r}} - \left(1-\frac{1}{\ep{r}\mu_{r}}\right) \#\left(\Vt \w \Mb{U}\right)  \\
    \Mh{U}  &\;\;\approx\;\; \frac{\Mb{U}}{\mu_{0}\mu_{r}} + \left(1 - \frac{1}{\ep{r}\mu_{r}}\right) \#\left(\Vt \w \Md{U}\right).\\
\end{split}
\end{eqnarray}

\section{Polarization and Magnetization}
The polarization 2-form $\Pi$ in spacetime is defined by
\begin{eqnarray}
    \Pi &=& G - \ep{0}F.\label{POL}
\end{eqnarray}
The second  macroscopic Maxwell equation may then be written
\begin{eqnarray*}
    \ep{0}d\star F &=& j - d\star\Pi = j + j_{p},
\end{eqnarray*}
where
\begin{eqnarray}\label{Polcurrent3}
    j_{p} &=& -d\star\Pi
\end{eqnarray}
will be called the electric polarization current 3-form. With
respect to {\it any} observer frame $U$ its orthogonal
decomposition is
\begin{eqnarray}\label{Pi}
    \Pi &=& \Mp{U} \w \wt{U} - \star\left(\frac{\Mm{U}}{\cc} \w \wt{U} \right) = \Mp{U} \w \wt{U} - \frac{1}{\cc}\MM{U} ,
\end{eqnarray}
where $\MM{U}=\#\Mm{U}$ and we call
\begin{eqnarray*}
    \Mp{U} = i_{U}\Pi \qquad \text{and} \qquad \frac{\Mm{U}}{\cc} = i_{U}\star \Pi
\end{eqnarray*}
the spatial polarization 1-form and magnetization 1-form
respectively relative to $U$. The Hodge dual of $\Pi$ has the
decomposition
\begin{eqnarray}\label{starPi}
    \star\Pi &=& \star(\Mp{U} \w \wt{U}) + \frac{\Mm{U}}{\cc} \w \wt{U}  = \MP{U} + \frac{1}{\cc}\Mm{U} \w \wt{U} ,
\end{eqnarray}
where $\MP{U}=\#\Mp{U}$. From (\ref{Media_G}),   (\ref{intro_F}),
(\ref{POL}) and  (\ref{Pi}) it follows
\begin{eqnarray}\label{coPolarCR}
    \Md{U} = \ep{0}\Me{U} + \Mp{U}  \qquad &\text{and}& \qquad \Mh{} = \mu_{0}^{-1}\Mb{U} + \Mm{U}.
\end{eqnarray}
From (\ref{Polcurrent3}), (\ref{Pi})   and (\ref{LIE}) one finds
\begin{eqnarray*}
    j_{p} &=& -d\MP{U} - \frac{d\Mm{U}}{\cc} \w \wt{U} = -\sd\MP{U} + \wt{U} \w \Lie_{U}\MP{U} - \frac{\sd\Mm{U}}{\cc} \w \wt{U}  \\
          &=& -\sd\MP{U} + \frac{1}{\cc}\left( \cc\Lie_{U}\MP{U} -\sd\Mm{U} \right) \w \wt{U}  \\
          &=& -\sd\MP{U} + \frac{1}{\cc}\left( \dotMP{U} - \sd\Mm{U} \right) \w \wt{U}.
\end{eqnarray*}
Writing the orthogonal decomposition of $j_{p}$ with respect to
$U$ as
\begin{eqnarray*}
    j_{p} &=& \frac{\J{p}}{\cc} \w \wt{U} + \RU{p},
\end{eqnarray*}
it follows that
\begin{eqnarray*}\label{polarJrho}
    \frac{\J{p}}{\cc} = -i_{U}j_{p} = \frac{1}{\cc}\left( \dotMP{U} - \sd\Mm{U}\right) \quad \text{and} \quad \RU{p} = -(i_{U}\star j_{p})\star\wt{U} = -\sd\MP{U}.
\end{eqnarray*}
In the frame $U$, $\J{p}$ and $\RU{p}$ denote the induced electric
polarization current density spatial 2-form and induced
polarization charge density spatial 3-form respectively. In a
similar manner
\begin{eqnarray*}
    d\Pi &=& d\Mp{U} \w \wt{U} - \frac{1}{\cc}d\MM{U} = \sd\Mp{U} \w \wt{U} - \frac{1}{\cc}\sd\MM{U} + \frac{1}{\cc}\wt{U} \w \Lie_{U}\MM{U} \\
         &=& \frac{1}{\cc}\left( \cc\sd\Mp{U} + \Lie_{U}\MM{U}\right) \w \wt{U} - \frac{1}{\cc}\sd\MM{U}.
\end{eqnarray*}
with the orthogonal decomposition $j_{m}=d\Pi=\frac{\J{m}}{\cc} \w
\wt{U} + \RU{m}$ where
\begin{eqnarray*}\label{MagJrho}
    \frac{\J{m}}{\cc} = -i_{U}j_{m} = \frac{1}{\cc}\left(\cc\sd\Mp{U} + \frac{1}{\cc}\dotMM{U}\!\right) \quad \text{and} \quad \RU{m} = -(i_{U}\star j_{m})\star\wt{U} = -\frac{1}{\cc}\sd\MM{U}
\end{eqnarray*}
denote the induced magnetization charge current density spatial
2-form and induced magnetization charge density spatial 3-form
respectively in terms of $\Mp{U}$ and $\MM{U}$.\\








\begin{thebibliography}{}
\bibitem{DGT} \BY{T. Dereli, J. Gratus \atque R. W. Tucker}
    \TITLE{The Covariant Description of Electromagnetically Polarizable Media}
    \IN{Phys. Lett. A}{361}{2006}{190-193}

\bibitem{DGT2} \BY{T. Dereli, J. Gratus \atque R.W. Tucker}
    \TITLE{New Perspectives on the Relevance of Gravitation for the Covariant Description of Electromagnetically Polarizable Media}
    \IN{J. Phys A: Math. Theor.}{10}{2007}{5695-5715}

\bibitem{PLAYER} \BY{M. A. Player}
    \TITLE{On the Dragging of the Plane of Polarization of Light Propagating in a Rotating Medium}
    \IN{Proc. Roy. Soc. A}{349}{1976}{441-445}

\bibitem{GBP} \BY{J. Gotte, S M. Barnett \atque M. Padgett}
    \TITLE{On the Dragging of Light by a Rotating Medium}
    \IN{Proc. Roy. Soc. A}{463}{2007}{2185-2194}

\bibitem{GCM8} \BY{R. W. Tucker \atque T. J. Walton}
    \TITLE{An Intrinsic Approach to Forces in Magnetoelectric Media}
    \IN{Il Nuovo Cimento C}{32}{2009}{205-229}
\end{thebibliography}
\end{document}